\documentclass[onecolumn,twoside]{article}

\usepackage{latexsym,amssymb,bm,bbold,amsfonts,amstext,graphicx,bbm,bm,relsize,dsfont}

\usepackage{authblk}
\usepackage{bold-extra}

\usepackage[table]{xcolor}
\usepackage{amsmath} 
\usepackage{amsthm} 
\usepackage{enumerate} 
\usepackage{float}
\usepackage[title]{appendix}
\usepackage{lipsum}
\usepackage{soul,color}

\usepackage[most,skins]{tcolorbox}
\usepackage{hyperref}
\hypersetup{
    unicode=true, 
    colorlinks=true, 
    linkcolor=blue, 
    citecolor=blue,
    urlcolor=blue
}

\usepackage[top=3cm, bottom=2.3cm, outer=2.5cm, inner=2.5cm, marginparwidth=2.5cm, marginparsep=0.5cm]{geometry}

\usepackage[utf8]{inputenc}
\usepackage[english]{babel}
\usepackage{enumitem}
\usepackage{media9}

\definecolor{comment}{HTML}{9326ff}

\definecolor{commentH}{HTML}{EE82EE}

\definecolor{commentJ}{HTML}{23ba4c}

\newcommand{\phdagger}{{\phantom{\dagger}}}

\newcommand{\tr}{\mathrm{Tr}}

\def\bra#1{\langle{#1}|}
\def\ket#1{|{#1}\rangle}
\def\braket#1{\langle{#1}\rangle}
\newcommand{\ketbra}[2]{\ket{#1}\!\bra{#2}}
\definecolor{lightterminal}{HTML}{f7f7ff}
\definecolor{energy_orange}{HTML}{ffdbc9}
\definecolor{light_blue}{HTML}{f2f5ff}

\definecolor{darkterminal}{HTML}{8792f5} 
\definecolor{quasiblack}{HTML}{545880} 
\definecolor{softblue}{HTML}{4759ff} %
\definecolor{deep_purple}{HTML}{6f00ff}
\definecolor{boxcolor}{HTML}{e5e3fa}

\usepackage{verbatim}
\usepackage[frozencache,cachedir=minted-cache]{minted}

\usepackage{etoolbox,xpatch}

\makeatletter
\AtBeginEnvironment{minted}{\dontdofcolorbox}
\def\dontdofcolorbox{\renewcommand\fcolorbox[4][]{##4}}
\xpatchcmd{\inputminted}{\minted@fvset}{\minted@fvset\dontdofcolorbox}{}{}
\xpatchcmd{\mintinline}{\minted@fvset}{\minted@fvset\dontdofcolorbox}{}{} 
\makeatother

\setminted[python]{xleftmargin=20pt,baselinestretch=1.2,linenos}
\setminted[MATLAB]{xleftmargin=20pt,baselinestretch=1.2,linenos}
\setminted[Mathematica]{xleftmargin=20pt,baselinestretch=1.2,linenos}

\newtcbox{\circled}[1][lightterminal]{on line,
    colback=#1, colframe=#1, boxsep=0pt, boxrule=0pt, size=small, arc=2pt}
    
\newtcolorbox[auto counter, number within = section, number freestyle={\noexpand\thesection.\noexpand\arabic{\tcbcounter}}]{code_box}[4]{breakable,center, fonttitle = \bfseries\sffamily, coltitle = white, colback = lightterminal, fontlower = \footnotesize, arc=1mm, width=\textwidth, frame style = softblue, pad at break=3mm, boxrule=0.5pt,
break at=-\baselineskip/0pt, enhanced jigsaw,
height fixed for=middle, before upper={\parindent7pt\noindent}, toprule=0pt, boxsep = 5pt, left=4pt, right=1.5pt,
title={\small\textcolor{energy_orange}{Script{ }\thetcbcounter:}{ } #2 {\hspace{5pt}} \href{#1}{\includegraphics[width=2.5mm]{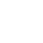}} \hfill \circled{\textcolor{quasiblack}{\texttt{#4}}}}, label=#3}

\newtcolorbox{eqbox}{center, breakable, colback = lightterminal, arc=2mm, colframe = softblue,  boxrule=0.8pt, width = \textwidth}

\newcommand{\code}[5]{
\begin{code_box}{#1}{#2}{#3}{#4}
\footnotesize
\inputminted{#4}{#5}
\end{code_box}
}

\newtcolorbox{codeline}[1]{colback=lightterminal, colframe=softblue,fonttitle=\bfseries,colbacktitle=softblue, enhanced, attach boxed title to top left={yshift=-7.1mm, xshift = 1.6mm},title={\footnotesize\texttt{#1}},boxrule=0.5pt}

\usepackage{titletoc}

\usepackage{fancyhdr,etoolbox}
\titlecontents{section}[1em]{\sffamily\bfseries}
{\thecontentslabel.\enspace}
{}
{\mdseries\titlerule*[0.5pc]{.}\quad\contentspage}[\vskip 4pt]

\titlecontents{subsection}[2.7em]{\sffamily}
{\thecontentslabel.\enspace}
{}
{\mdseries\titlerule*[0.5pc]{.}\quad\contentspage}[\vskip 3pt]

\titlecontents{subsubsection}[5em]{\sffamily}
{\thecontentslabel.\enspace}
{}
{\mdseries\titlerule*[0.5pc]{.}\quad\contentspage}[\vskip 3pt]

\usepackage{etoolbox}
\pretocmd{\contentsname}{\sffamily}{}{}

\usepackage{titlesec, blindtext, color}

\titleformat{\section}[hang]{\Large\sffamily\bfseries}{{\thesection.}{ }}{0pt}{\Large\bfseries}

\titleformat{\subsection}[hang]{\large\sffamily\bfseries}{{\thesubsection.}{ }}{0pt}{\large\bfseries}

\titleformat{\subsubsection}[hang]{\sffamily\bfseries}{{\thesubsubsection.}{ }}{0pt}{\bfseries}

\usepackage{abstract}

\usepackage[labelfont={bf,sf}]{caption}

\usepackage{csquotes}

\usepackage{times}

\makeatletter
\renewcommand{\maketitle}{\bgroup\setlength{\parindent}{0pt}
\begin{flushleft}
  \textbf{\@title}

  \@author
\end{flushleft}\egroup
}
\makeatother

\renewenvironment{abstract}
 {\par\noindent\textsf{\textbf{\abstractname}}\ \ignorespaces\vspace{10pt} \\}
 {\par\medskip}

\title{\sffamily{\bfseries{\LARGE{A Tutorial on Quantum Master Equations:\vspace{5pt} \\ {\Large Tips and tricks for quantum optics, quantum computing and beyond}}}}\vspace{10pt}}

\makeatletter
\def\@xfootnote[#1]{%
  \protected@xdef\@thefnmark{#1}%
  \@footnotemark\@footnotetext}
\makeatother

\author[1,2]{Francesco Campaioli\footnote[$\dagger$]{\href{francesco.campaioli@rmit.edu.au}{francesco.campaioli@rmit.edu.au}}}

\author[1]{Jared H. Cole\footnote[*]{\href{jared.cole@rmit.edu.au}{jared.cole@rmit.edu.au}}}

\author[1]{Harini Hapuarachchi\footnote[$\ddagger$]{\href{harini.hapuarachchi@rmit.edu.au}{harini.hapuarachchi@rmit.edu.au}}}

\affil[1]{\small{\textit{Chemical and Quantum Physics, and ARC Centre of Excellence in Exciton
Science,}
\textit{School of Science, RMIT University, Melbourne 3000, Australia}}}

\affil[2]{\small{\textit{Dipartimento di Fisica e Astronomia “G. Galilei,” Università degli Studi di Padova, I-35131 Padua, Italy,}

\textit{Padua Quantum Technologies Research Center, Università degli Studi di Padova, I-35131 Padua, Italy}}
}

\begin{document}

\date{}

\maketitle

\thispagestyle{empty}

\begin{abstract}
Quantum master equations are an invaluable tool to model the dynamics of a plethora of microscopic systems, ranging from quantum optics and quantum information processing, to energy and charge transport, electronic and nuclear spin resonance, photochemistry, and more. This tutorial offers a concise and pedagogical introduction to quantum master equations, accessible to a broad, cross-disciplinary audience. The reader is guided through the basics of quantum dynamics with hands-on examples that build up in complexity. The tutorial covers essential methods like the Lindblad master equation, Redfield relaxation, and Floquet theory, as well as techniques like Suzuki-Trotter expansion and numerical approaches for sparse solvers. These methods are illustrated with code snippets implemented in \texttt{python} and other languages, which can be used as a starting point for generalisation and more sophisticated implementations. 
\end{abstract}

\hypersetup{linkcolor = black}
\tableofcontents
\hypersetup{linkcolor = blue}

\section{Introduction}
\label{s:introduction}

\thispagestyle{empty}

Master equations are differential equations used to model the dynamics of systems that can be described as a probabilistic combination of some states. For example, the concentration dynamics of a chemical reaction $ x\rightleftharpoons y$, where some reactants $x$ lead to some products $y$, can be described by the differential equations,
\begin{equation}
    \label{eq:synthesis_example}
    \begin{cases}
    &\dot{p}_x = k_{y\to x} p_{y} -k_{x\to y} p_x , \\
    &\dot{p}_y = k_{x\to y} p_x - k_{y\to x} p_{y},
    \end{cases}
\end{equation}
where $p_i$ represent the concentrations of species $i=x,y$, with $\dot{p}_i = dp_i/dt$ being their time derivative, and $k_{i\to j}$ the transition rates from species $i$ to $j$. This equation can be easily solved to obtain the transient and steady state concentration of the reactants and products, as a function of their initial concentrations and transition rates. In a reaction like the one modelled in
Eq.~\eqref{eq:synthesis_example}, the total concentration is conserved, since $\dot{p}_\mathrm{tot} := \dot{p}_x+\dot{p}_y = 0$. Then, by recasting the problem in terms of relative concentrations $p_i \to p_i/p_\mathrm{tot}$, we can interpret $p_i$ as \textit{the probability of being in state $i$}. We can generalise this idea to formulate master equations as first-order differential equations to the vector of probabilities $\bm{p} = (p_1,\cdots,p_n)$ of being in one of the $n$ states of some system of interest. As a result, the dynamics of the states probabilities are prescribed by the master equation
\begin{equation}
    \label{eq:general_master_equation}
    \dot{\bm{p}} = \bm{F}(\bm{p},t),
\end{equation}
with $\bm{F}$ often being a linear function of $\bm{p}$ represented by some generating matrix $A$, as in $\dot{\bm{p}} = A \bm{p}$.

However, when dealing with quantum systems we must take into account that coherent superpositions of states participate in the evolution, as prescribed by Schr\"odinger's equation
\begin{equation}
    \label{eq:schroedinger}
    \frac{d}{dt}\ket{\psi(t)} = -\frac{i}{\hbar} H\ket{\psi(t)},
\end{equation}
where $H$ is the Hamiltonian of the system, and $\ket{\psi(t)} = \sum_{j=1}^n c_j(t) \ket{\phi_j}$ is its state at time $t$, expressed as a coherent superposition of the eigenstates $\mathcal{B}_H = \{\ket{\phi_i},\cdots,\ket{\phi_n}\}$ of the Hamiltonian, via the normalised complex coefficients $c_j(t)$ satisfying $\sum_i |c_i(t)|^2 = 1$.
In this case, a vector of probabilities $\bm{p}$, with $p_i = |c_i|^2$, is no longer sufficient to completely describe the dynamics of the system, since different phases of $c_i$ will lead to different solutions. Master equations for the dynamics of quantum systems can then be expressed by employing another representation of the state of the system, known as the density operator $\rho$. As discussed in details in Sec.~\ref{s:density_operators}, the density operator contains all the information regarding the probabilities (known as \textit{populations}) of being in each state $i$, given by $p_i = \braket{\phi_i|\rho|\phi_i}$, as well as the phases (known as \textit{coherences}) $\varphi_{ij} = \braket{\phi_i|\rho|\phi_j}$ associated with the coherent superpositions between basis states $\ket{\phi_i}$ and $\ket{\phi_j}$. Quantum master equations are then formulated by generalisation of Eq.~\eqref{eq:general_master_equation}, as first-order differential equations to the density operator,
\begin{equation}
    \label{eq:general_quantum_master_equation}
    \dot{\rho} = \mathcal{F}(\rho,t).
\end{equation}

In this tutorial we will primarily cover a specific type of linear quantum master equations (QMEs), that respect a set of requirements for the evolution of the density operator, as discussed in Sec.~\ref{s:doma}. QMEs, initially developed in quantum optics to study light-matter interactions~\cite{Carmichael1999}, have been adopted in a multitude of settings, across different disciplines and fields, such as photochemistry~\cite{Atkins1974,Iwasaki2001,Forecast2023}, energy and charge 
transport~\cite{Plenio2008,Mohseni2008,Lee2015}, high-precision magnetometry~\cite{Betzholz2014,Jeske2017,Hapuarachchi2022},
electronic~\cite{Nakano2016,Norambuena2020,Kobori2020,Collins2022} and nuclear spin resonance~\cite{Redfield1955,Hendrickson1973,Jeener1998}, quantum information processing~\cite{Sarandy2005,Verstraete2009,Keck2017,Campaioli2022}, thermodynamics in the quantum regime~\cite{Uzdin2015,Farina2019,Gherardini2020,}, and are certainly not limited to these settings. One of they key aspects of QMEs is that they provide a coarse-grained stochastic description of the effect of unknown and uncontrollable agents on a system of interest~\cite{Breuer2002}, leading to a computationally inexpensive ensemble-averaged picture of the dynamics of quantum systems. QMEs can be phenomenological~\cite{Genkin2008} or derived, using first principles~\cite{Breuer2002}, from a microscopic model of the system-environment interactions, as done in Sec.~\ref{s:bloch-redfield}. They can be used to derive qualitative trends~\cite{Albers1971} or make quantitatively accurate predictions~\cite{Hapuarachchi2022}. They are just as suitable for the derivation of analytical results~\cite{Atkins1974} as they are for the numerical simulation of complex systems with a large number of degrees of freedom~\cite{Campaioli2021}. For these reasons, QMEs have become a standard approach to model the dynamics of quantum systems, and a starting point for the formulation of more sophisticated descriptions.

Quantum master equations are now more accessible than ever, thanks to the many dedicated libraries and software packages, such as \href{https://qutip.org/}{\texttt{QuTiP}}~\cite{Johansson2012}, \href{https://github.com/USCqserver/HOQSTTutorials.jl}{\texttt{HOQST}}~\cite{Chen2022},  \href{https://spindynamics.org}{\texttt{Spinach}}~\cite{Hogben2011}, and \href{https://github.com/jevonlongdell/qotoolbox}{\texttt{qotoolbox}}, to name a few. These resources offer an invaluable platform for the quick implementation of models and their systematic exploration. Indeed, they have established themselves as a staple tool on the workbench of a vast community of researchers. Pedagogical tutorials and documentations of these libraries are just as precious as the software itself, offering an accessible starting point and a pathway for rapid progression. Nevertheless, when directing newcomers from different research areas to QMEs, an obstacle is often presented by the vast and technical library of resources like textbooks and notes, written for a specialised audience, which may not be ideal for cross-disciplinary readers. To bridge this gap, this tutorial provides the reader with a concise introduction to quantum master equations, with a pedagogical, hands-on approach, in the style of an interactive lesson or a workshop. The aim is to provide a handbook for third-year students joining the research group, master students ready to implement models, and PhD students and cross-disciplinary researchers looking to consolidate and expand their expertise. 

In this tutorial we cover essential theories, like the Lindblad master equation, Bloch-Redfield theory and Floquet theory, as well as numerical techniques for their solutions, such as the stochastic wavefunction method, the Suzuki-Trotter expansion, and numerical approaches for sparse matrices.
We illustrate these methods using scripts implemented in \texttt{python}. Building up in complexity, these examples aim to provide a deeper understanding of the methods implemented behind the curtains in libraries like \texttt{QuTiP} and \texttt{qotoolbox}, and can be used as a starting point for generalisations. 
Versions of these scripts in \texttt{MATLAB} and \texttt{Mathematica} can be found in Appendix~\ref{a:examples}. 

\section{Density Operators} 
\label{s:density_operators}
In this section we briefly review the mathematical description of the \textit{state} of a quantum system, focusing on the numerical implementation of state vectors and density operators. We assume that the reader is familiar with the \textit{postulates of Quantum Mechanics}, Hilbert spaces, expectation values, time evolution, and composite systems, which can be reviewed in any of these textbooks~\cite{Cohen1978,Scully1997,Gardiner2000,Bransden2000,Breuer2002,Schlosshauer2007,Wiseman2009,Weiss2012,,Nielsen2010}. 

\subsection{Pure states}
\label{ss:pure_states}
Let us consider a $d$-dimensional quantum system with Hilbert space $\mathcal{H}$. Let $\mathcal{B}:= \{ |\phi_1\rangle, |\phi_2\rangle, ..., |\phi_d\rangle\}$ be an orthonormal basis for $\mathcal{H}$, so that $\langle \phi_i|\phi_j\rangle = \delta_{ij}$. For example, $\mathcal{B}$ could be given by the orthonormal eigenstates of a hermitian operator such as some Hamiltonian $H$. 
Any state of the system can be expressed as a \emph{coherent superposition} with complex coefficients $c_i\in\mathbb{C}$,
\begin{equation}
\label{eq:coherent_superposition}
	|\psi\rangle = c_1|\phi_1\rangle+c_2|\phi_2\rangle+...+ c_d|\phi_d\rangle = \sum_{j=1}^d c_j\ket{\phi_j},
\end{equation}
where the coefficient $c_j$ are such that $\braket{\psi|\psi} = \sum_{j=1}^d |c_j|^2= 1$, according to the Born interpretation of the wavefunction~\cite{Mcmahon2013}.
The square of the coefficients in Eq.~\eqref{eq:coherent_superposition}, $|c_j|^2$, represents the probability of finding the system in the eigenstate $|\phi_j\rangle$ upon measurement in the considered basis $\mathcal{B}$. See Ref.~\cite{Cohen1978} for a review of projective measurement and Ref.~\cite{Nielsen2010} for the generalisation to positive operator valued measures (POVMs). 

Unit vectors like $\ket{\psi}$ are called \emph{pure states}. A pure state contains all the available physical information about the system, such as the expectation value of an observable $\mathcal{A}$ associated with hermitian operator $A$,
\begin{equation}
    \label{eq:pure_expectation_value}
    \braket{A} = \braket{\psi|A|\psi}.
\end{equation}
The following \texttt{python} script uses methods from the \href{https://numpy.org/}{\texttt{numpy}} library to implement state vectors and operators, and calculates the expectation value of some observable.
\code{https://github.com/frnq/qme/blob/main/python/pure_states_expect.py}{Pure states and expectation values}{code:pure_states_expect}{python}{Scripts/python/pure_states_expect.txt}

\subsection{Mixed states: Proper and improper mixtures}
\label{ss:mixed_states}

There are two important scenarios where pure states are no longer sufficient to describe the state of a system. First, in experimental settings, we often lack the knowledge of the exact pure state $|\psi\rangle$ of our system. Instead, we may know that the system is in any of the pure orthonormal states $\{\ket{\psi_j}\}$ with some probabilities $\{ p_j \}$. In other words, our knowledge of the system is represented by a \emph{statistical mixture of pure states}, described by the set $\lbrace|\psi_j\rangle, p_j\rbrace$. 
In such case, when more than one $p_j$ is non-zero, the system is said to be in a \emph{mixed state}. This is sometimes referred to as a \textit{proper} mixture~\cite{Masillo2009}.

Second, when studying the dynamics of composite systems, pure states are no longer the most general description of a state. This is because the marginal state of any \textit{entangled} state cannot be represented as a pure state, and instead, needs to be represented as a statistical mixture over the basis elements of the considered subsystem~\cite{Nielsen2010}, as discussed in Sec.~\ref{ss:composite systems}. This is sometimes referred to as an \textit{improper mixture}~\cite{Masillo2009}. See Refs.~\cite{Cohen1978,Nielsen2010} for more on composite systems, and Refs.~\cite{Mintert2005,Bengtsson2006,Modi2012} for an in-depth analysis of entanglement and other quantum correlations. 

\subsection{Definition and properties of the density operator}
\label{ss:properties_density_operator}
Whether we are dealing with proper or improper mixtures of states, we can represent the set $\{\ket{\psi_j},p_j\}$ using a linear operator on the Hilbert space,
\begin{equation}
\label{eq:density_operator}
	{\rho} = \sum_{j=1}^{d}p_j \ketbra{\psi_j}{\psi_j},
\end{equation}
known as the density operator~\cite{Nielsen2010}, where $\ketbra{\psi_j}{\psi_j}$ is the outer product of $\ket{\psi_j}$ with itself, that is, the vector product of $\ket{\psi_j}$ with its dual $\bra{\psi_j}$. The coefficients $p_j > 0$ are such that $\sum_j p_j = 1$, since they represent probabilities (also known as \textit{convex combination}).  Density operators have three fundamental properties,
\begin{enumerate}
    \item \textbf{Hermitian:} ${\rho} = {\rho}^\dagger$. This implies that $\rho$ has only real eigenvalues.
    \item \textbf{Positive\footnote{Or, more specifically, \textit{positive semi-definite}.}:} $\rho > 0$. That is, $\rho$ eigenvalues $p_j\in[0,1]$ are not negative.
    \item \textbf{Normalised:} $\tr\rho= 1$, which can also be stated as $\sum_j p_j = 1$, i.e., the sum of its eigenvalues (probabilities) must add up to 1.
\end{enumerate}

Density operator can represent both pure and mixed states, and can be expressed in any basis $\mathcal{B} = \{\ket{\phi_i}\}_{i=1}^d$ of the Hilbert space $\mathcal{H}$ as
\begin{equation}\label{Eq:Density_Matrix_in_basis}
	{\rho} = \sum_{i,j=1}^{d}\rho_{ij}\ketbra{\phi_i}{\phi_j} = \begin{pmatrix}
	\rho_{11} & \rho_{12} &\dots &\rho_{1d}\\
	\rho_{21} & \rho_{22} &\dots &\rho_{2d}\\
	\vdots & \vdots & \ddots & \vdots\\
	\rho_{d1} & \rho_{d2} &\dots &\rho_{dd}
	\end{pmatrix},
\end{equation}
where $\rho_{ij}$ is the associated matrix element with row $i$ and column $j$.
The diagonal elements $\rho_{ii}$ of the density matrix are known as \emph{populations} and they denote the probabilities of finding the system in the respective basis states $|\phi_i\rangle$. The off-diagonal elements $\rho_{ij}$ are known as \emph{coherences}, and provide information about the coherent superposition of the basis states $|\phi_i\rangle$ and $|\phi_j\rangle$~\cite{Manzano2020}. 

Similarly to state vectors, density operators encode all the available information that can be extracted from the considered system. For example, the expectation value of some observable $\mathcal{A}$ associated with hermitian operator $A$ can be calculated as,
\begin{equation}
    \label{eq:expectation_mixed}
    \braket{A} = \tr[A\rho].
\end{equation}
The following \texttt{python} script provides an implementation of a density operator and the evaluation of the expectation value of some observable. There, a system with dimension $d=3$ is in a mixed state defined by state vectors $\{\ket{\psi_1},\ket{\psi_2},\ket{\psi_3}\}$ with probabilities $\{0.1,0.3,0.6\}$, represented by the density operator $\rho = 0.1\ketbra{\psi_1}{\psi_1}+0.3\ketbra{\psi_2}{\psi_2}+0.6\ketbra{\psi_3}{\psi_3}$.
\code{https://github.com/frnq/qme/blob/main/python/mixed_states_expect.py}{Mixed states and expectation values}{code:mixed_states_expect}{python}{Scripts/python/mixed_states_expect.txt}

\subsection{Composite systems}
\label{ss:composite systems}

Composite systems consist of two or more (interacting) quantum systems, whose Hilbert space is given by the tensor product of the individual Hilbert subspaces, $\mathcal{H} = \bigotimes_{i} \mathcal{H}_i$~\cite{Nielsen2010}. For example, a composite system might be given by a pair of interacting two-level systems (\textit{qubits}, in quantum information theory), or by a system $\mathrm{S}$ interacting with some large environment $\mathrm{E}$.

\subsubsection{Tensor product and partial trace}
\label{sss:tensor_product_partial}
Any state $\rho$ of a composite system can be represented using a basis $\mathcal{B}$ constructed using the \textit{tensor product} of the basis elements of each subsystems' basis $\mathcal{B}_\alpha = \{ \ket{\phi_i}_\alpha \}_{i=1}^{d_\alpha}$. For example, a bipartite system can be expressed in the following basis,
\begin{equation}
    \mathcal{B} = \Big\{ \ket{\phi_i}_1\otimes\ket{\phi_j}_2 \Big\}_{i,j}.
\end{equation} 
In \texttt{python}, the tensor product can be implemented with \texttt{numpy} using the Kroneker product \href{https://numpy.org/doc/stable/reference/generated/numpy.kron.html}{\texttt{kron}}. 
\begin{codeline}{python}
\footnotesize
    \begin{minted}[linenos=false]{python}
        psi = numpy.kron(psi1,psi2)
    \end{minted}
\end{codeline}
\noindent
Similar implementations are available in \texttt{Mathematica} and \texttt{MATLAB}, with \href{https://reference.wolfram.com/language/ref/KroneckerProduct.html}{\texttt{KroneckerProduct}} and \href{https://au.mathworks.com/help/matlab/ref/kron.html}{\texttt{kron}}, respectively.

When taking expectation values for composite systems, it may be useful to focus only on the \textit{marginal state} of one of the subsystems. For example, the marginal state $\rho_1$ of subsystem $1$ is obtained from the total state $\rho$ by \textit{tracing over} the degrees of freedom associated with the rest of the Hilbert space (here, subsystem $2$),
\begin{equation}
    \label{eq:partial_trace}
    \rho_1 = \tr_2 [\rho].
\end{equation}
The linear operator $\tr_i[\cdot]$ is called \textit{partial trace}, and its definition can be found in Ref.~\cite{Breuer2002}. For the case of bipartite systems with dimensions $d_1$ and $d_2$, the partial trace can be implemented in \texttt{python} using \texttt{numpy}.
\begin{codeline}{python}
\footnotesize
    \begin{minted}[linenos=false]{python}
        rho1 = np.trace(rho.reshape(d1,d2,d1,d2), axis1=0, axis2=2)
        rho2 = np.trace(rho.reshape(d1,d2,d1,d2), axis1=1, axis2=3)
    \end{minted}
\end{codeline}

For example, let us consider the following bipartite pure state 
\begin{equation}
    \label{eq:bipartite_pure}
    \ket{\psi(\theta)} = \cos(\theta)\ket{00}+\sin(\theta)\ket{11},
\end{equation}
where $\ket{00}=\ket{0}_1\otimes\ket{0}_2$, $\ket{11}=\ket{1}_1\otimes\ket{1}_2$, and its associated density operator is given by $\rho(\theta) = \ketbra{\psi(\theta)}{\psi(\theta)}$. The state $\rho(\theta)$ is separable for $\theta = 0,\pi/2$, and entangled otherwise, being maximally entangled\footnote{The state $\ket{\psi(\pi/4)}$ is the $\Phi_+$ Bell state~\cite{Nielsen2010}.} for $\theta = \pi/4$. 
As a result, for $\theta\neq k\pi/2$ the partial state of each subsystem $\rho_i(\theta) = \tr_j[\rho(\theta)]$ is not pure, and is therefore an improper mixture.

To measure the degree of mixedness of a density operator we can use the \textit{purity} $\mathcal{P}$,
\begin{equation}
    \label{eq:purity}
    \mathcal{P}[\rho] = \tr[\rho^2] = \sum_{j=1}^d p_j^2,
\end{equation}
which is bounded between 1, for pure states $\rho = \ketbra{\psi}{\psi}$, and $1/d$, for maximally mixed states $\rho = \mathbb{1}/d$. For more on purity, entropy, measures of distinguishability, and other information-theoretic figures of merit see Refs.~\cite{Bengtsson2006,Nielsen2010}.

\begin{figure}[h]
    \centering
    \includegraphics{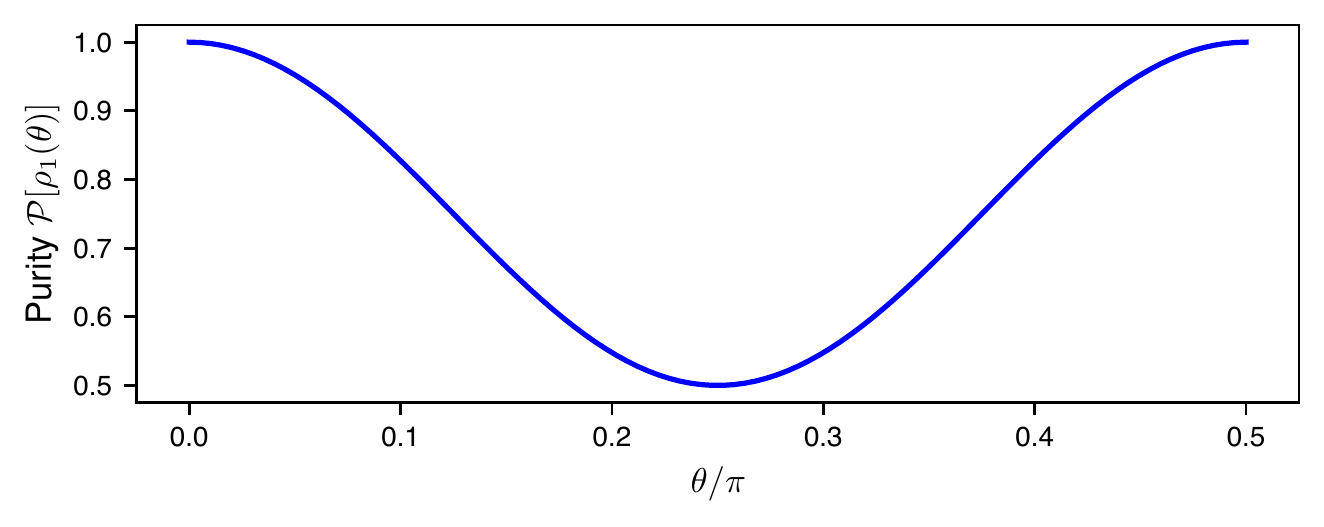}
    \caption{Purity of the marginal state $\rho_1(\theta) = \tr_2\rho(\theta)$, calculated using script~\ref{code:partial_trace}. The state $\rho_1(\theta)$ is maximally mixed for $\theta = \pi/4$, since $\rho(\pi/4)$ is maximally entangled. This is an example of an improper mixture.}
    \label{fig:partial_trace}
\end{figure}

The following \texttt{python} script calculates the marginal state of the first subsystem, $\rho_1(\theta) = \tr_2\rho(\theta)$, showing that its purity $\mathcal{P}[\rho_1(\theta)]<1$ for $\theta \neq k\pi/2$. Notice that $\rho_1(\theta)$ is maximally mixed when $\rho(\theta)$ is maximally entangled, i.e., $\tr\rho_1(\pi/4)=1/2$, as shown in Fig.~\ref{fig:partial_trace}. A powerful implementation of the tensor product and the partial trace (\href{https://qutip.org/docs/3.1.0/guide/guide-tensor.html}{\texttt{ptrace}}) for any type of composite system is available in \href{https://qutip.org/}{\texttt{QuTiP}}, as shown in the script~\ref{code:tensor_partial}.
\code{https://github.com/frnq/qme/blob/main/python/partial_trace.py}{Partial trace and purity of entangled states}{code:partial_trace}{python}{Scripts/python/partial_trace.txt}

\subsubsection{Direct sum}
\label{sss:direct_sum}

Sometimes, it is useful to compose systems given by the \textit{addition} of different Hilbert spaces together. For example, when studying a pair of interacting systems with Hilbert space $\mathcal{H}_a = \mathcal{H}_1\otimes\mathcal{H}_2$ and dimension $d_a$, it might be convenient to add some states $\{\ket{\phi_i}_b\}_{i=1}^{d_b}$ to the picture, perhaps representing the result of some transitions that are modelled phenomenologically.
In these cases the total Hilbert space is given by
\begin{equation}
    \label{eq:direct_sum}
    \mathcal{H} = \mathcal{H}_a \oplus \mathcal{H}_b.
\end{equation} 
Numerically, a basis for this space can be constructed, from the bases of each individual subsystem, using a block matrix structure,
\begin{equation}
    M = \begin{pmatrix} 
        M_a & \bm{0} \\
        \bm{0}^\mathrm{T} & M_b
    \end{pmatrix},
\end{equation}
where $M_a$ and $M_b$ are $d_a \times d_a$ and $d_b \times d_b$ matrices, respectively, and $\bm{0}$ is a $d_a\times d_b$ matrix. The above structure can be implemented in \texttt{python} using the following script. For more information on tensor products, direct sums, and irreducible representations, see Ref.~\cite{Cohen1978}.

\code{https://github.com/frnq/qme/blob/main/python/block_matrix.py}{Composing a block-matrix operator}{code:block_matrix}{python}{Scripts/python/block_matrix.txt}

\subsection{Schr\"odinger and von Neumann equations}
\label{ss:schrodinger_von_neumann}

When studying the dynamics of quantum systems using the density operator representation, Schr\"odinger's equation~\eqref{eq:schroedinger} becomes,
\begin{equation}
    \label{eq:von_neumann_equation}
    \dot{\rho}(t) = - \frac{i}{\hbar}[H,\rho(t)], 
\end{equation}
known as the \textit{von Neumann}\footnote{Or \textit{Liouville-von Neumann} equation.} equation, where $H$ is the Hamiltonian of the system (which can be time-dependent), $\dot{\rho} = \partial_t\rho$, and $[\cdot,\cdot]$ is the commutator~\cite{Breuer2002}. In general, the solution to this equation is given by some unitary operator $U(t;t_0)$ that propagates the state of the system from some initial time $t_0$ to some time $t$,
\begin{equation}
    \label{eq:solution}
    \rho(t) = U(t;t_0) \rho(t_0) U(t;t_0)^\dagger,
\end{equation}
where $\dagger$ is the conjugate transpose (\textit{adjoint)}. If $H$ is time-independent the solution is given by $U(t;t_0) = \exp[-i H (t-t_0)/\hbar]$ and can be reduced to $U(\tau) = \exp[-i H \tau/\hbar]$ for all $t,t_0$ such that $\tau = t-t_0$. See Ref.~\cite{Joachain1975,Breuer2002} for more on the solution $U$ for time-dependent Hamiltonian using time-ordering operators and the Dyson series.

\subsubsection{Open quantum systems}
\label{sss:open_quantum_systems}
The focus of this tutorial is the dynamics of systems that interact with their surrounding environment. These can be seen as composed of a system of interest $\mathrm{S}$ and an environment $\mathrm{E}$ that is usually large, uncontrollable, or not experimentally accessible~\cite{Breuer2002}. The dynamics of the full composite system $\mathrm{S}$-$\mathrm{E}$ (or \textit{universe}) follows equation Eq.~\eqref{eq:von_neumann_equation} with Hamiltonian
\begin{equation}
    \label{eq:open_system_hamiltonian}
    H = H_\mathrm{S} + H_\mathrm{E} + H_\mathrm{int},
\end{equation}
where $H_\mathrm{int}$ represents the interaction between the system with Hamiltonian $H_\mathrm{S}$ and the environment with Hamiltonian $H_\mathrm{E}$.

If the solution $U(t;t_0)$ is known, the dynamics of the system $\mathrm{S}$ can be drawn from the state of the universe $\rho$ by tracing over the environment's degrees of freedom,
\begin{equation}
    \rho_\mathrm{S}(t) = \tr_\mathrm{E} \big[ \rho(t) \big].
\end{equation}
However, finding $U$ for large composite systems is often a difficult problem, both numerically and analytically. Instead, we may seek to obtain a prescription for the dynamics of the system's state by performing the partial trace of Eq.~\eqref{eq:von_neumann_equation}, to obtain
\begin{equation}
\label{eq:reduced_von_neumann_equation}
    \dot{\rho}_\mathrm{S}(t) = -\frac{i}{\hbar}\tr_\mathrm{E}\big\{ [H,\rho(t)] \big\}.
\end{equation}
Eq.~\eqref{eq:reduced_von_neumann_equation} provides the starting point for the derivation of density operator master equations such as those reviewed in Secs.~\ref{s:doma} and~\ref{s:bloch-redfield}.

\section{Density operator master equations}
\label{s:doma}

Density operator master equations are a powerful tool to study the dynamics of quantum systems that interact weakly with their surrounding environment. Originally developed in the field of quantum optics to study light-matter interactions~\cite{Carmichael1999}, they are used to simulate a variety of quantum mechanical phenomena, such as noise models for quantum information processing~\cite{Nielsen2010}, transient emission and absorption spectra of optically active materials~\cite{Buchheit2016}, and electronic and nuclear spin resonance experiments~\cite{Goldfarb2018}.

The power of master equations resides in the choice of ignoring the environment's dynamics, often uncontrollable and inaccessible. By neglecting the environment's degrees of freedom, we can limit the scaling of the computational requirements to a polynomial of $d = \mathrm{dim}\mathcal{H}_\mathrm{S}$, where $\mathcal{H}_\mathrm{S}$ is the system's Hilbert space. In this section we introduce quantum master equations and focus on their numerical implementation and solution, providing direction for further readings. 

\subsection{Introduction to Lindblad master equation}
The paradigmatic example of a density operator master equation is the Gorini-Kossakowski-Sudarshan-Lindblad (GKSL) master equation~\cite{Milz2017}, often known as the \textit{Lindblad} master equation,
\begin{eqbox}
    \begin{equation}
        \label{eq:lindblad_master_equation}
        	\dot{{\rho}}(t) = -\frac{i}{\hbar}[H, {\rho(t)}] + \sum_{k}\gamma_k\bigg(L^\phdagger_k{\rho(t)}L_k^\dagger -\frac{1}{2}\Big\{ L_k^\dagger L_k^\phdagger, {\rho(t)} \Big\}\bigg),
    \end{equation}
\end{eqbox}
\noindent
where $\rho$ is the system's density operator\footnote{From now on we will drop the subscript $\mathrm{S}$ from the system's density operator, unless specified otherwise.}, $H$ is the system Hamiltonian, and $\{L_k\}$ are the \textit{Lindblad} operators\footnote{Also known as \textit{collapse} operators or \textit{jump} operators.} representing some non-unitary processes like relaxation or decoherence that occur at some rates $\{\gamma_k\}$. The operators $[.,.]$ and $\lbrace.,.\rbrace$ denote the commutator and anti-commutator of the operands. Note that, from now on $H$ will represent the system's Hamiltonian, unless specified otherwise.

Like the Hamiltonian generates coherent dynamics, the Lindblad operators\footnote{Formally, the Lindblad operators are dimensionless linear combinations of the basis operators in Liouville space~\cite{Breuer2002}, and therefore the index $k$ in the sum of Eq.~\eqref{eq:lindblad_master_equation} can be limited to $d^2-1$.} generate incoherent transitions in the space of states. Unlike the Hamiltonian, they do not need to be hermitian. For example, a decay transition from some excited state $\ket{e}$ to some ground state $\ket{g}$ is mediated by the Lindblad operator
\begin{equation}
    \label{eq:lindblad_op_example}
    L_\downarrow = \ketbra{g}{e}.
\end{equation}
Indeed, when we apply $L_\downarrow$ to $\ket{e}$ we obtain $\ket{g} = L_\downarrow\ket{e}$. Note that $L_\downarrow^\dagger = \ketbra{e}{g} \neq L_\downarrow$.

Eq.~\eqref{eq:lindblad_master_equation} is used to approximate the evolution of the density operator of a system $\mathrm{S}$ with Hamiltonian $H$ that is weakly coupled to a Markovian (memory-less) environment~\cite{Breuer2002}. The Lindblad master equation is the general form for a completely positive and trace-preserving (CPTP) Markovian and time-homogeneous map for the evolution of the system's density operator $\rho$~\cite{Breuer2002}. More on the motivation for the requirements of CPTP and Markovianity can be found in Refs.~\cite{Breuer2002,Milz2017}. Derivations of Eq.~\eqref{eq:lindblad_master_equation} can be found in Refs.~\cite{Breuer2002,Lidar2019,Manzano2020}. 

\subsection{The Liouville superoperator}
\label{ss:superoperator}

When solving Eq.~\eqref{eq:lindblad_master_equation}, it is convenient to express the master equation in a vector notation, 
\begin{equation}
\label{eq:superop_form}
\dot{\bm{\rho}} = \mathcal{L} \bm{\rho},
\end{equation}
known as \textit{superoperator} or \textit{Liouville} form, 
where $\bm{\rho} = \mathrm{vec}({\rho})$ is the \textit{vectorised} form of ${\rho}$, and $\mathcal{L}$ is the superoperator associated with the generator $\dot{\rho}$ of Eq.~\eqref{eq:lindblad_master_equation}. The matrix associated with the density operator ${\rho}$ can be reshaped into a vector in many equivalent ways\footnote{Column and row ordering are common choices.} resulting in different superoperators. Any reshaping is valid, as long as one keeps track of the ordering in the elements of the superoperator. The following is a \texttt{python} script that illustrates a reshaping via the \texttt{numpy} method \texttt{reshape}:

\code{https://github.com/frnq/qme/blob/main/python/vectorising_rho.py}{Vectorising a density matrix}{code:vectorising_rho}{python}{Scripts/python/vectorising_rho.txt}
\noindent 
Similar methods are available in \texttt{Mathematica} and \texttt{MATLAB}. A robust implementation of the reshaping is implemented in \texttt{QuTiP} with the methods \texttt{operator\_to\_vector} and \texttt{vector\_to\_operator}.

\subsubsection{Constructing the Liouville superoperator}
\label{sss:constructing_superoperator}
While the superoperator $\mathcal{L}$ can be constructed ``by hand'' for small systems, it is advisable to have a systematic approach to compile it from some Hamiltonian $H$ and some Lindblad operators $\{L_k\}$. Two common ways are to either follow an index prescription for the superoperator tensor $\rho_{ab} = \sum_{cd}\mathcal{L}_{abcd}\rho_{cd}$, or to use the following linear algebra identity for the column-ordered form of $\mathrm{vec}(\rho)$~\cite{Barnett1990, Byron2012}:
\begin{equation}
\label{eq:row_ordered_identity}
	\mathrm{vec}({A}{X}{B}) = ({B}^\mathrm{T}\otimes{A})\mathrm{vec}({X}).
\end{equation}
To take advantage of the latter, we proceed inserting the identity operator $\mathbb{1}$ into Eq.~\eqref{eq:lindblad_master_equation}
\begin{equation}
\label{eq:lindblad_with_idenity}
\mathbb{1}\dot{{\rho}}\mathbb{1} = -\frac{i}{\hbar}\Big(H{\rho}\mathbb{1} - \mathbb{1}{\rho}H\Big) + \sum_k\gamma_k\bigg(L_k^\phdagger{\rho}L_k^\dagger -\frac{1}{2}\Big(L_k^\dagger L_k^\phdagger{\rho}\mathbb{1} + \mathbb{1}{\rho}L_k^\dagger L_k^\phdagger\Big)\bigg),
\end{equation}
from which the superoperator can be easily constructed using the tensor product structure discussed in Sec.~\ref{ss:composite systems}, and implemented with the \texttt{kron} method in \texttt{python} and \texttt{MATLAB} or the  \texttt{KroneckerProduct} function in \texttt{Mathematica}. Combining Eqs.~\eqref{eq:superop_form},~\eqref{eq:row_ordered_identity} and~\eqref{eq:lindblad_with_idenity}
we obtain
\begin{equation}
\label{eq:liouville_ordered}
\mathcal{L} = -\frac{i}{\hbar}\Big(\mathbb{1} \otimes H - H^\mathrm{T}\otimes\mathbb{1}\Big) + \sum_k\gamma_k \bigg( L_k^*\otimes L_k^\phdagger - \frac{1}{2} \Big(\mathbb{1}\otimes L_k^\dagger L_k^\phdagger + L_k^\mathrm{T} L_k^*\otimes\mathbb{1}\Big)\bigg).
\end{equation}
As discussed in the next sections, the power and advantage of the superoperator form consists in offering a direct pathway to solving Eq.~\eqref{eq:lindblad_master_equation}, based on the solution of a system of linear ordinary differential equations. The following \texttt{python} script implements Eq.~\eqref{eq:liouville_ordered} using \texttt{numpy} arrays. It is worth noting that the rates $\gamma_k$ are here embedded into the Lindblad operators via $L_k \to L'_k = \sqrt{\gamma_k} L_k$ for a simpler implementation. 
\code{https://github.com/frnq/qme/blob/main/python/superoperator.py}{Constructing the superoperator}{code:superoperator_python}{python}{Scripts/python/superoperator.txt}
\noindent 

\subsection{Steady-state solution}
\label{ss:staeady_state}

Before looking at the dynamics $\rho(t)$ of the density operator, let us go through some methods to obtain the steady-state solution of Eqs.~\eqref{eq:lindblad_master_equation} and~\eqref{eq:superop_form}.

\subsubsection{Using the null space of Liouville superoperator}
Once we have expressed a linear master equation in the superoperator form, we can use the matrix $\mathcal{L}$ to study the behaviour of the system. Of immediate interest is the steady state solution ($\dot{{\rho}} = 0$) which is often measured directly in experiments. To find any steady state solutions we solve for the \textit{null space} of $\mathcal{L}$~\cite{Axler1997}, which is the subspace of all vectors $\bm{\rho}$ that satisfy the equation
\begin{equation}
\label{eq:null_space_equation}
	\mathcal{L}\bm{\rho}=0.
\end{equation}
Numerically, this can be done using the \texttt{NullSpace} function in \texttt{Mathematica}, the \texttt{null} function in \texttt{MATLAB}, or the \texttt{null\_space} method in the \texttt{numpy} library \texttt{scipy}. An analytic solution can also be sought with this approach with \texttt{Mathematica}, or \texttt{SymPy} in \texttt{python}. If there is a unique solution, solving for the null space will provide the corresponding steady state density matrix vector $\bm{\rho}(\infty)$ up to a constant factor, the value of which is given by the original normalization condition $\text{Tr}({\rho})=1$.
 
If there are multiple solutions, solving for the null space will give linearly independent vectors. In such case, the steady state depends on the initial state of the system. For example, let us consider a two-level system Hamiltonian\footnote{In Sec.~\ref{ss:two_level_atom_with_field} we outline how to obtain (\ref{eq:tl_hamiltonian}) for a two-level atom interacting with an electric field. } $H$ with energy splitting $\Delta$ and coupling $\Omega$, and Lindblad operators $L_\downarrow$ and $L_0$ associated with spontaneous relaxation and dephasing, respectively,
\begin{equation}
\label{eq:tl_hamiltonian}
	{H} = \hbar\begin{pmatrix}
	0 & \Omega\\
	\Omega & \Delta
	\end{pmatrix}, \;\;\;\ L_\downarrow = \sqrt{\gamma_\downarrow} \begin{pmatrix}
    0 & 1\\
    0 & 0
    \end{pmatrix}, 
    \;\;\;\ L_0 = \sqrt{\gamma_0} \begin{pmatrix}
    1 & 0\\
    0 & 1
    \end{pmatrix},
\end{equation}
where $\gamma_\downarrow$ and $\gamma_0$ are the rates associated with relaxation and dephasing.
In the following script, we construct $\mathcal{L}$ in \texttt{python} and solve for its null space for the case of (i) relaxation and no-driving limit $\Omega,\gamma_0=0$, and (ii) dephasing and driving, with no relaxation, $\gamma_\downarrow = 0$.

\code{https://github.com/frnq/qme/blob/main/python/null_space.py}{Solving for the null space of superoperator {\normalfont\textsf{(requires script~\ref*{code:superoperator_python})}}}{code:null_space}{python}{Scripts/python/null_space.txt}
\noindent
In the limit of relaxation and no-driving, there is a unique steady state $\bm{\rho}(\infty) = \begin{pmatrix} 1,0,0,0 \end{pmatrix}^\textrm{T}$, which is the ground state of the system, as expected for a two-state system undergoing spontaneous relaxation with no driving field. Instead, for the case of dephasing and driving, the \texttt{null} function returns two vectors that span the two-dimensional linear subspace associated with the null space of $\mathcal{L}$. In this case, the specific steady state depends on the choice of initial state. See Sec.~\ref{a:examples} for a \texttt{MATLAB} implementation of the method used in script~\ref{code:null_space}.

\subsubsection{Algebraic solution}
The steady state solution ${\rho}(\infty)$ for both linear and non-linear\footnote{A non-linear generator is such that $\mathcal{L}(\rho)$ depends on the state of the system. See Ref.~\cite{Breuer2002} for more on non-linear density operator master equations.} generators can be obtained by solving Eq.~\eqref{eq:lindblad_master_equation} for $\dot{{\rho}} = 0$ algebraically (or symbolically). In \texttt{python}, this can be done using the \texttt{solve} method of the \texttt{SymPy} library, as demonstrated in the script below for the case of $\Omega=0, \gamma_0 = 0$, with respect to Eq.~\eqref{eq:tl_hamiltonian}.
\code{https://github.com/frnq/qme/blob/main/python/symbolic_solution.py}{Symbolic steady-state solution}{code:symbolic_matrix_solving_python}{python}{Scripts/python/symbolic_solution.txt}

\noindent
Algebraic solutions can also be sought in \texttt{MATLAB} with \texttt{solve}, or in \texttt{Mathematica}, using the \texttt{Solve} method. These provide a more straightforward approach to solving symbolic matrix equations. A \texttt{MATLAB} implementation of script~\ref{code:symbolic_matrix_solving_python} can be found in the Appendix in script~\ref{code:symbolic_matrix_solving}. 

\subsection{Solving the dynamics of the system}
\label{ss:solving_dynamics}

\begin{figure}
    \centering
    \includegraphics{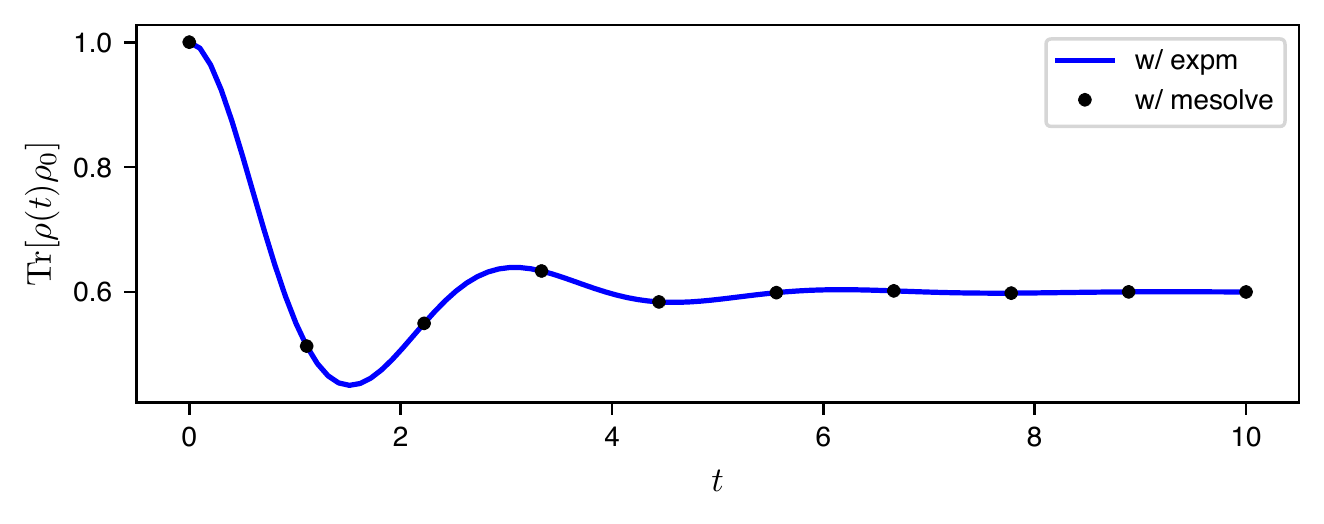}
    \caption{Propagation using matrix exponential (\texttt{expm} from \texttt{numpy}), obtained using script~\ref{code:super_matrix_exp}. The propagated state $\rho(t)$ is obtained using Eq.~\eqref{eq:matrix_exp_superoperator}, and compared to the solution obtained using a finite-difference method (\texttt{mesolve} from \texttt{QuTiP}), implemented in script~\ref{code:mesolve_example}.}
    \label{fig:super_matrix_exp}
\end{figure}

Let us now discuss how to solve Eq.~\eqref{eq:lindblad_master_equation} in order to obtain the state of the system $\rho(t)$ at any time $t$ from a given initial condition $\rho_0 = \rho(t_0)$. Let us represent the solution with the dynamical map $\rho(t) = \Lambda(t;t_0)[\rho_0]$. For linear, time-independent generators $\mathcal{L}$, the solution to Eq.~\eqref{eq:superop_form} can be obtained by calculating the following matrix exponential~\cite{Breuer2002},
\begin{equation}
    \label{eq:matrix_exp_superoperator}
    \bm{\rho}(t) = \exp\big[\mathcal{L} (t-t_0)\big]\bm{\rho}(t_0).
\end{equation}
The operator $P(t;t_0) = \exp\big[\mathcal{L} (t-t_0)\big]$ is called the propagator of the evolution. From the propagator, we can obtain the solution $\rho(t)$ by reshaping $\bm{\rho}(t)$ as described earlier in this section. See Ref.~\cite{Lidar2019} for details on how to obtain the dynamical map $\Lambda$ from $P$ using, for example, a Kraus operators representation. Eq.~\eqref{eq:matrix_exp_superoperator} is implemented in \texttt{python} using \texttt{scipy} in the following script, with the result shown in Fig.~\ref{fig:super_matrix_exp}.
\code{https://github.com/frnq/qme/blob/main/python/super_matrix_exp.py}{Propagator using matrix exponential {\normalfont\textsf{(requires script~\ref*{code:superoperator_python})}} }{code:super_matrix_exp}{python}{Scripts/python/super_matrix_exp.txt}
\noindent
See script~\ref{code:solution_matlab_ode} for an implementation of Eq.~\eqref{eq:matrix_exp_superoperator} using \texttt{MATLAB}.

It is worth pointing out that the approach used in script~\ref{code:super_matrix_exp} is by no mean optimised, and calculates a new propagator for every time step in the considered time domain. When working with evenly-spaced time steps we can reduce the computational cost by exploiting the composition rule of dynamical semigroups, as discussed in Sec.~\ref{ss:semigroup}. In \texttt{QuTiP}, instead, the solution is obtained using the sophisticated and powerful \texttt{mesolve} method, which by default uses \texttt{scipy}'s numerical integration library \href{https://docs.scipy.org/doc/scipy/tutorial/integrate.html}{\texttt{integrate}}. 

\subsubsection{Singular value decomposition of the Liouville superoperator}\label{sss:td:superop}

The superoperator $\mathcal{L}$ is generally a complex, non-Hermitian matrix. For this reason a spectral decomposition of $\mathcal{L}$ is not always guaranteed, that is, $\mathcal{L}$ may not admit the diagonal representation $\mathcal{L} = V D V^{-1}$. However, $\mathcal{L}$ always admits a singular value decomposition\footnote{The generalisation of the eigenvalue decomposition.} (SVD), and therefore can be represented in terms of its left and right-singular vectors, $\bm{L}_k$ and $\bm{R}_k$, respectively, and the set of complex singular values $\{\lambda_k\}$, that abide by the following relationships~\cite{Sacha2020},
\begin{align}
    \mathcal{L} \bm{R}_k &= \lambda_k \bm{R}_k, \label{eq:right_eig_vec}\\
    \bm{L}^\dagger_k \mathcal{L} &= \lambda_k \bm{L}^\dagger_k.  
    \label{eq:left_eig_vec}
\end{align}
Notice that both $\bm{R}_k$ and $\bm{L}_k$ are column vectors, hence $\bm{L}^\dagger_k$ is a row vector. Each left and right-singular vectors can be normalized via,
\begin{align}
    \hat{\bm{R}}_k &= \bm{R}_k\big/\sqrt{\bm{L}^\dagger_k\bm{R}_k}
    \label{eq:Norm_r}\\
    \hat{\bm{L}}^\dagger_k &= \bm{L}^\dagger_k\big/\sqrt{\bm{L}^\dagger_k\bm{R}_k}.
    \label{eq:norm_Ldag}
\end{align}
The normalized singular vector pairs then follow the usual orthonormalisation condition~\cite{Sacha2020},
\begin{equation}
\label{eq:bi-orthonormality}
    \hat{\bm{L}}^\dagger_i  \hat{\bm{R}}_j^\phdagger = \delta_{ij}. 
\end{equation}
 
The solution of Eq.~\eqref{eq:superop_form} for a system with time independent Liouville superoperator $\mathcal{L}$ can now be expressed as follows,
\begin{equation}
\label{eq:solution_svd}
    \bm{\rho}(t) = \sum_{k=1}^{d^2}  \hat{\bm{L}}^\dagger_k\bm{\rho}(t_0) \hat{\bm{R}}_k e^{\lambda_k (t-t_0)}
\end{equation}
where $d$ is the dimension of the Hilbert space. 

The advantage of expressing the time evolution in form of Eqs.~\eqref{eq:matrix_exp_superoperator} and~\eqref{eq:solution_svd} is that it is exact (when the singular values are found exactly) for all times and therefore does not depend on the step size or other operational details of the integration routine used to solve the differential equation. In the following \texttt{python} code, we use \texttt{scipy} to obtain the temporal solutions for a system with,
\begin{equation}
\label{eq:singular_value_problem}
	H = \hbar\begin{pmatrix}
	0 & \Omega\\
	\Omega & 0
	\end{pmatrix},
	\quad \quad
    L = \begin{pmatrix}
    0 & 1\\
    1 & 0
    \end{pmatrix}, \quad \textrm{and}\quad
    \bm{\rho}(0) = \begin{pmatrix}0, 0, 0, 1\end{pmatrix}^\textrm{T}
\end{equation}
using the singular value decomposition of $\mathcal{L}$.

\code{https://github.com/frnq/qme/blob/main/python/dynamics_norm_svd.py}{Solution using normalized singular vectors {\normalfont\textsf{(requires script~\ref*{code:superoperator_python})}}}{code:temporal_python_solution}{python}{Scripts/python/dynamics_norm_svd.txt}

\noindent
The solution is shown in Fig.~\ref{fig:temporal_python_solution}. A \texttt{MATLAB} implementation of this method can be found in the Appendix in script~\ref{code:temporal_solution}.

\begin{figure}
    \centering
\includegraphics{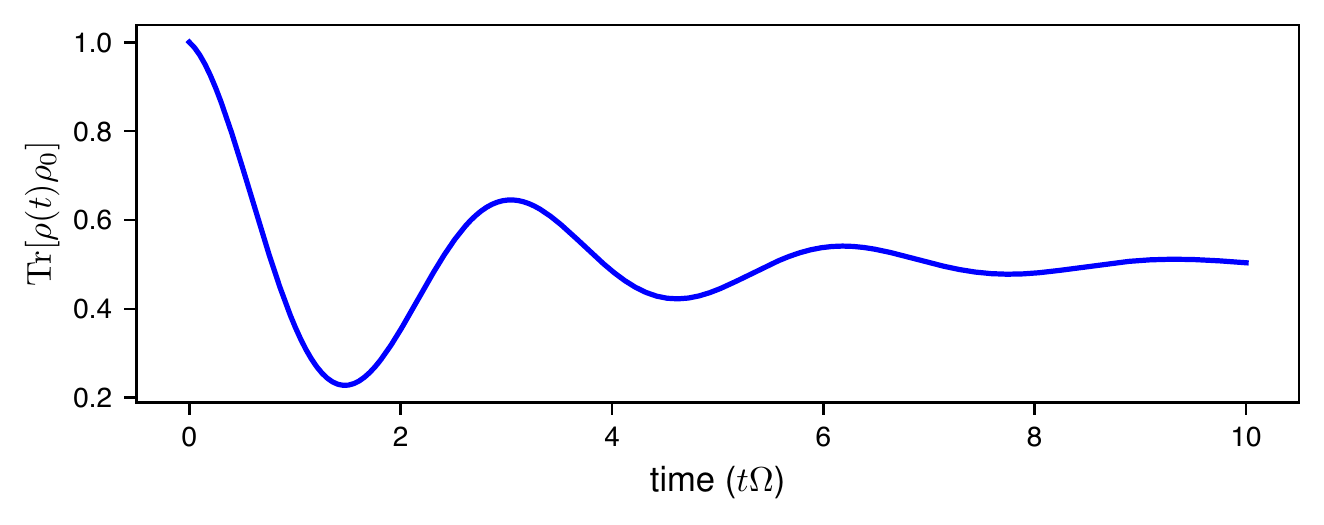}
    \caption{Dynamics of the state $\rho(t)$ for Eq.~\eqref{eq:singular_value_problem}, solving the Lindblad master equation using the normalised superoperator singular vectors, obtained with script~\ref{code:temporal_python_solution}. }
    \label{fig:temporal_python_solution}
\end{figure}

\subsubsection{Time-dependent generators}
\label{sss:time-dependent_generators}
If the Hamiltonian or the decoherence terms depend on time, Eqs.~\eqref{eq:superop_form} is generalised to
\begin{equation}
    \label{eq:time-dependent_generator}
    \dot{\bm{\rho}} = \mathcal{L}(t) \bm{\rho},
\end{equation}
where the Liouville generator $\mathcal{L}(t)$ now explicitly depends on time. In this case the solution of Eq.~\eqref{eq:matrix_exp_superoperator} is not valid. The general solution of Eq.~\eqref{eq:time-dependent_generator} is given by
\begin{equation}
    \label{eq:time-ordered_solution}
    \bm{\rho}(t) = \mathcal{T}\big\{ \exp[\textstyle\int_0^t ds \mathcal{L}(s)] \big\} \bm{\rho}(t_0),
\end{equation}
where $\mathcal{T}$ is the time-ordering operator, analogue to the Dyson series for time-dependent Hamiltonians and wavefunction propagation~\cite{Breuer2002,Schnell2020}. Eq.~\eqref{eq:time-ordered_solution} can be approximated, for instance, by means of a sequence of step-wise time-independent generators, before resorting to other means like numerical integration. 

If the generator $\mathcal{L}(t)$ is approximately \emph{piecewise time-independent}, then Eq.~\eqref{eq:matrix_exp_superoperator} can be applied to each time slice, using the result of the previous slice to provide the input state for the next slice. This scenario is common in many optical and spin resonance experiments. For example, it can be used to
compute the effect of applying a laser pulse resonant with an atomic transition, to then observing the behaviour of the system while the pulse is on and immediately after it has been turned off. 

\begin{figure}
    \centering
    \includegraphics{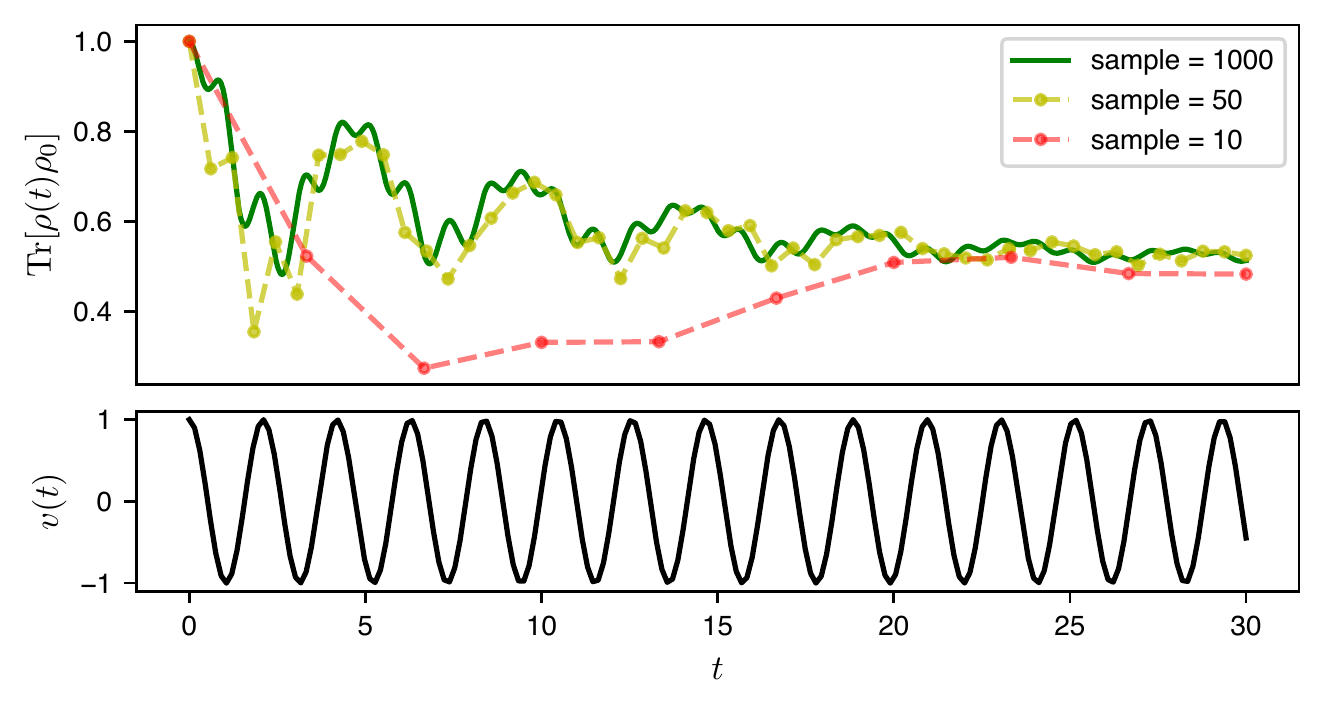}
    \caption{Solution of time-dependent generator using piecewise time-independent propagator, for the system considered in  script~\ref{code:time_dependent_generator}. The evolution is generated by a time-dependent Hamiltonian $H(t) = \omega_0\sigma_z/2 + v(t)\omega_0\sigma_x/2$, with $v(t)=\cos(\omega t)$, and a time-independent Lindblad dephasing operator $J = (\mathbb{1}-\sigma_z)/2$, associated with rate $\gamma$. The solution is obtained for $\omega_0 = 1$, $\omega =  3$ and $\gamma = 0.3$.}
    \label{fig:td_propagation}
\end{figure}

For example, let us consider a system with Hamiltonian $H = H_0 + v(t) H_1$, where $H_0 = \omega_0\sigma_z/2$, $H_1 = \omega_0\sigma_x/2$, $v(t) = \cos(\omega t)$, and a Lindblad dephasing operator $J = \ketbra{g}{g} =(\mathbb{1}-\sigma_z)/2$, with dephasing rate $\gamma$. The generator $\mathcal{L}(t) = \mathcal{L}_0+\mathcal{L}_1(t)$ can be split into a time-independent part $\mathcal{L}_0$, associated with $H_0$ and $L_0$, and a time-dependent part $\mathcal{L}_1(t)$. To reduce the computational cost when propagating this system, we can update the propagator by updating only the time-dependent part.
The following \texttt{python} script generalises the solution of Eq.~\eqref{eq:matrix_exp_superoperator} to the case of time-dependent generators, by updating the superoperator at each time $t$; the solution is shown in Fig.~\ref{fig:td_propagation}. Note that for this approach to be accurate, the time step $\delta t$ has to be sufficiently small so that $v(t+\delta t)\approx v(t)+\mathcal{O}(\delta t^2)$. For rapidly varying time-dependent Hamiltonians other methods are required. If $H(t)$ is periodic, a solution can be found using an effective time-independent Hamiltonian, obtained using Floquet theory, as discussed in Sec.~\ref{s:oscillatory}.

\code{https://github.com/frnq/qme/blob/main/python/time_dependent_generator.py}{Solution of time-dependent generator {\normalfont\textsf{(requires script~\ref*{code:superoperator_python})}}}{code:time_dependent_generator}{python}{Scripts/python/time_dependent_generator.txt}

Note that in practice, especially when using theoretical system parameters, it is often possible to get exact cancellations which may have no physical grounding but can result in degenerate eigenvectors. While there are mathematical techniques which deal with these situations, it is often easier to just add an infinitesimal (numerically of order machine precision) imaginary term $i\varepsilon$, $\varepsilon\ll1$, to each element of the matrix. This can remove the degeneracy, even if the term is made suﬃciently small to have no perceivable effect on the resulting calculations.

\subsubsection{Propagation via semigroup composition}
\label{ss:semigroup}

The dynamical maps generated by a linear Markovian quantum master equation like Eq.~\eqref{eq:lindblad_master_equation} are a family of single-parameter maps $\Lambda_t$ that have the following composition property,
\begin{equation}
    \label{eq:quantum_dynamical_semigroup}
    \Lambda_s \circ \Lambda_t = \Lambda_{s+t}, \;\;\;\; t,s \geq 0,
\end{equation}
and, therefore, are known as a \textit{quantum dynamical semigroup} (QDS).
The above can also be expressed as $\Lambda_s[\Lambda_t[\rho]]=\Lambda_{s+t}[\rho]$. For more on QDS see Ref.~\cite{Breuer2002}. Eq.~\eqref{eq:quantum_dynamical_semigroup} can be expressed in the superoperator form as
\begin{equation}
    \label{eq:superoperator_semigroup}
    P(s)P(t) = P(s+t), \;\;\;\; t,s \geq 0,
\end{equation}
which follows directly from the properties of the exponential and the fact that $[\mathcal{L}s,\mathcal{L}t]=0$. Note that the above does not generally hold for time-dependent $\mathcal{L}(t)$ and non-linear generators $\mathcal{L}(\rho(t))$.

\begin{figure}[h]
    \centering
    \includegraphics{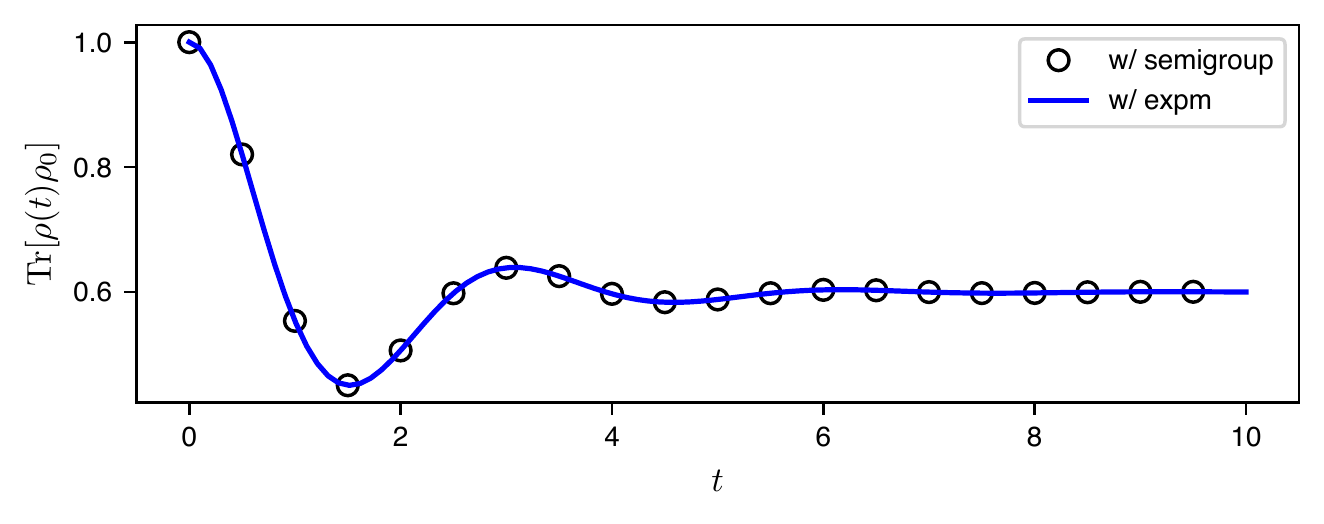}
    \caption{Propagation using semigroup decomposition, obtained using script~\ref{code:semigroup_propagator}, compared to that obtained by computing a new propagator for each time-step.}
    \label{fig:propagation_semigroup}
\end{figure}

When propagating a system in time over an evenly-spaced time set $\{k\delta t\}_{k=1}^{m}$ we can exploit the composition rule of dynamical semigroups to vastly reduce the computational cost of propagation. Instead of calculating a new propagator $P(t_k)$ for each time step $t_k = t_0 +k\delta t$, we can calculate a single propagator $P_1 = P(\delta t)$ and obtain all the others using
\begin{equation}
    \label{eq:propagator_semigroup}
    P(t_k) = \prod_{j=1}^{k} P_1 = P_1^k.
\end{equation}
The following \texttt{python} script implements Eq.~\eqref{eq:propagator_semigroup}, and the results are shown in Fig.~\ref{fig:propagation_semigroup}.
\code{https://github.com/frnq/qme/blob/main/python/semigroup_propagator.py}{Propagation using dynamical semigroup composition {\normalfont\textsf{(requires script~\ref*{code:superoperator_python})}}}{code:semigroup_propagator}{python}{Scripts/python/semigroup_propagator.txt}

This approach is particularly useful when propagating for very long times or when using large time-steps, in which cases \texttt{scipy}'s \texttt{integrate} methods usually tend to accumulate large numerical errors. When propagating over several orders of magnitude, it may be convenient to break each timescale into evenly-spaced time sets to resolve the details of different dynamical transients. For example, this is useful when looking at dynamics from the femtosecond to the nanosecond timescales.

\subsubsection{Baker–Campbell–Hausdorff \& Zassenhaus formula}
\label{ss:zassenhaus_BCH}

Hamiltonians and superoperators are often sums of two or more terms, such as $W = U+V$. As briefly noted in Sec.~\ref{ss:semigroup}, when the terms commute with each other $[U,V]=0$, the solution can be obtained from the composition of individual terms. For example, let $\mathcal{L} = \mathcal{L}_1 + \mathcal{L}_2$, with $[\mathcal{L}_1 ,\mathcal{L}_2] = 0$, then
\begin{equation}
    \label{eq:composition_commuting}
    P(t) = \exp(\mathcal{L} t) = \exp(\mathcal{L}_1 t)\exp(\mathcal{L}_2 t).
\end{equation}
Instead, when considering pairs of non-commuting operators $[X,Y]\neq 0$, we have $\exp(X+Y)\neq\exp(X)\exp(Y) = \exp(Z)$. The solution to the latter equation for $Z$ is known as the Baker-Campbell-Hausdorff (BCH) formula~\cite{Rossmann2002}, and reads,
\begin{equation}
    \label{eq:BCH}
    Z = X + Y + \frac{1}{2}[X,Y] + \frac{1}{12}\bigg( [X,[X,Y]] + [Y,[Y,X]]\bigg) + \cdots.
\end{equation}
The BCH solution finds application when used in the Zassenhaus formula, which allows us to decompose a matrix exponential $\exp[(X+Y)t]$, where $t$ is a scalar parameter, in terms of a product series,
\begin{equation}
    \label{eq:zassenhaus}
    \exp[(X+Y)t] = \exp[Xt]\exp[Yt]\exp\bigg[-\frac{1}{2}[X,Y]t^2\bigg]\exp\bigg[\frac{1}{3}\Big([Y,[X,Y]]+\frac{1}{2}[X,[X,Y]]\Big)t^3\bigg]\cdots.
\end{equation}
The formula becomes useful when the product series can be truncated or approximated to a certain set of terms. This is for example particularly useful when the generator is time-dependent $\mathcal{L}_t$ and $[\mathcal{L}_t,\mathcal{L}_s]\neq 0$: By choosing a sufficiently small time step $\delta t$ such that $s = t+\delta t$, the series of Eq.~\eqref{eq:zassenhaus} can be truncated to terms in $\mathcal{O}(\delta t^m)$ for some $m>1$, as discussed in the next section.

\subsubsection{Suzuki-Trotter expansion}
\label{ss:suzuki-trotter}

\begin{figure}[h]
    \centering
    \includegraphics{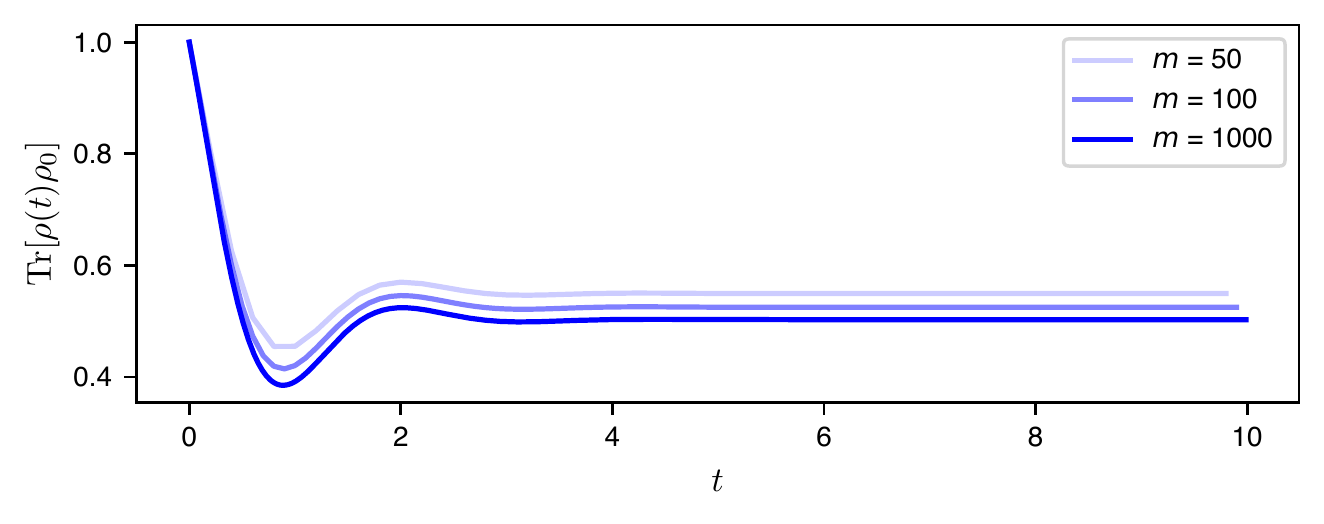}
    \caption{Propagation using Suzuki-Trotter expansion for different amounts $m = 50,100,1000$ of time steps, obtained using script~\ref{code:suzuki}.}
    \label{fig:suzuki_trotter}
\end{figure}

A consequence of the Zassenhaus formula is that, for \textit{small} time steps $\delta t$, Eq.~\eqref{eq:zassenhaus} can be truncated to the first order in $\delta t$ with errors of the order of $O(\delta t^2)$
\begin{equation}
    \label{eq:limit_zassenhaus}
    \exp[(X+Y)\delta t] = \exp[X\delta t]\exp[Y\delta t] + O(\delta t^2).
\end{equation}
This can be used to obtain the solution for long times using the product series,
\begin{equation}
    \exp[(X+Y)\delta t] = \lim_{n \to \infty} \bigg[ \exp\bigg(X \frac{t}{n}\bigg)\exp\bigg(Y\frac{t}{n}\bigg) \bigg]^n,
\end{equation}
also known as \textit{Suzuki–Trotter expansion} or \textit{Lie product formula}~\cite{Rossmann2002}. This approach is particularly useful when studying the dynamics of interacting many body systems or time-dependent generators. The following \texttt{python} script uses the Suzuki–Trotter expansion to propagate a system by separating the contribution of the two non-commuting superoperators. The results are shown in Fig.~\ref{fig:suzuki_trotter}.
\code{https://github.com/frnq/qme/blob/main/python/suzuki.py}{Propagation using Suzuki-Trotter expansion {\normalfont\textsf{(requires script~\ref*{code:superoperator_python})}}}{code:suzuki}{python}{Scripts/python/suzuki.txt}

\subsubsection{Numerical solution with finite-difference methods}

While the matrix exponential is a powerful tool to obtain exact solution of Eq.~\eqref{eq:superop_form}, it may be less computationally expensive to compromise some precision in favor of less demanding time and memory requirements. Not only finite-difference methods can prove efficient at solving density operator master equations, but they can also be used to solve the dynamics of non-linear and time-dependent generators.
In this case, the approach consists in solving the set of coupled differential equations obtained by element-wise comparison of the left and right hand sides of Eq.~\eqref{eq:lindblad_master_equation}. 

\begin{figure}[h]
    \centering
    \includegraphics{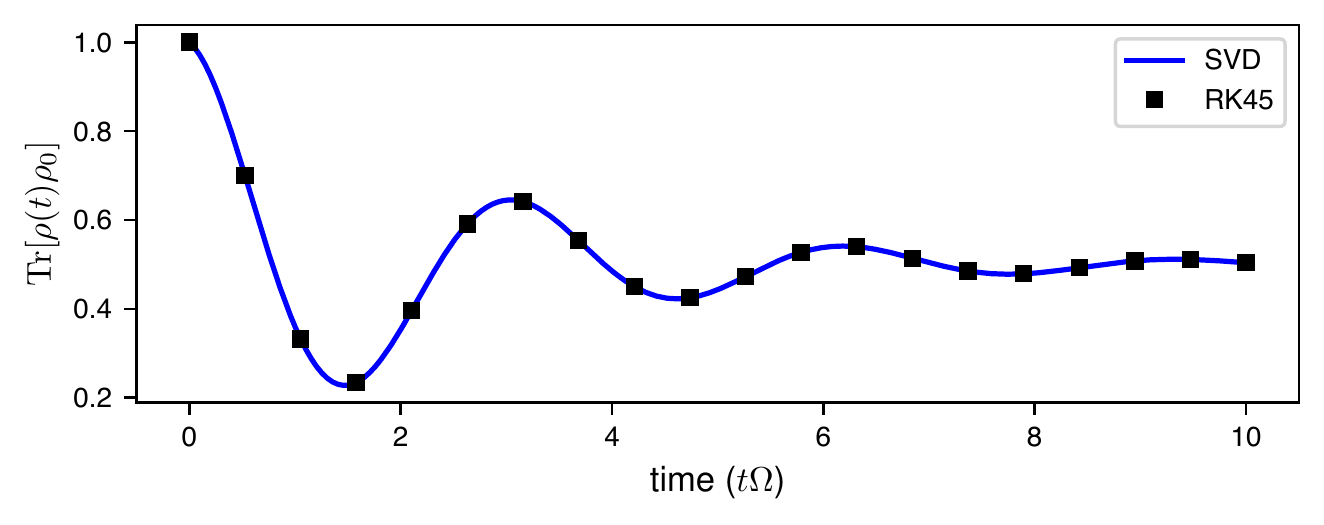}
    \caption{Propagation using finite-difference approach, based on the 4-5th order Runge-Kutta method. The solution is obtained using script~\ref{code:solution_python_ode}, and compared to that obtained using script~\ref{code:temporal_python_solution}, based on the singular value decomposition.}
    \label{fig:solution_rk45}
\end{figure}

The following script is a continuation of script~\ref{code:temporal_python_solution}, and solves the dynamics of the same two-level system using the 4-5th order Runge-Kutta differential equation method. The method is implemented using the initial-value problem solver \texttt{solve\_ivp} from \texttt{scipy.integrate} library for \texttt{python}. The solution is shown in Fig.~\ref{fig:solution_rk45}. A \texttt{MATLAB} implementation of the same code can be found in script~\ref{code:solution_matlab_ode} in the Appendix.

\code{https://github.com/frnq/qme/blob/main/python/dynamics_finite_diff.py}{Solution with finite-difference method {\normalfont \textsf{(requires script~\ref*{code:temporal_python_solution})}}}{code:solution_python_ode}{python}{Scripts/python/dynamics_finite_diff.txt}

\subsubsection{Solution using the stochastic wavefunction method}
\label{ss:swfm}

Since the amount of complex floating point numbers required to represent superoperators like $\mathcal{L}$ and $P$ scales as $d^4$, memory may become an issue for large systems. To circumvent this problem we can propagate a density operator using the stochastic wavefunction method~\cite{Carmichael1999}, also known as \textit{Monte Carlo wavefunction} method or \textit{master equation unravelling}. Originally developed for quantum optics, the method is an adaptation of the kinetic Monte Carlo method~\cite{Bortz1975} to the solution of Eq.~\eqref{eq:lindblad_master_equation}. 

Instead of propagating a density operator solving Eq.~\eqref{eq:superop_form}, the method provides a procedure to propagate a state vector $\ket{\psi_0}$ under the influence of some generator $\mathcal{L}$, by sampling a sufficiently large amount $N$ of stochastic trajectories $\Psi_j = \{\ket{\psi_j(t)}\}$, to then obtain the time-evolved density operator $\rho(t)$ by averaging over them,
\begin{equation}
    \label{eq:MCWF_rho_average}
    \rho(t) = \sum_{j=1}^{N} \ketbra{\psi_j(t)}{\psi_j(t)}.
\end{equation}

Let $H$ be the Hamiltonian of the system, and $\{L_k\}_{k=1}^{M}$ a collection of Hermitian\footnote{The method can be implemented with non-Hermitian Lindblad operators too, upon some adaptations to avoid division-by-zero errors in the normalisation steps.} Lindblad operators. In the simplest form of the method, each trajectory $\Psi_j$ is sampled according to the following steps:
\begin{enumerate}
    \item The probabilities associated with any of the $k$ incoherent transitions mediated by the $L_k$ jump operators is calculated, 
    \begin{equation}
        \label{eq:MCWF_prob_array}
        \delta p_k = \delta t \braket{\psi(t)|L_k^\dagger L_k^\phdagger|\psi(t)}\geq0,
    \end{equation}
    with $\delta p = \sum_{k=1}^{M}\delta p_k$.
    \item A uniform random number $u\in(0,1]$ is sampled. 
        \begin{enumerate}
            \item If $\delta p < u$, then no jump occurs and the state $\ket{\psi(t)}$ at time $t$ is evolved by means of the non-Hermitian effective Hamiltonian $H_\textrm{eff} = H - i\hbar \sum_{k=1}^M L_k^\dagger L_k^\phdagger /2$,
            \begin{equation}
                \label{eq:MCWF_non-hermitian_prop}
                \ket{\widetilde{\psi}(t+\delta t)} = \bigg(1-\frac{i}{\hbar} H_\textrm{eff}^\dagger \delta t\bigg)\ket{\psi(t)},
            \end{equation}
            where $\ket{\widetilde{\psi}}$ indicates that the state vector may not be normalised. 
            \item If $\delta p\geq u$, a jump occurs. A new uniform random number $u'\in(0,1]$ is sampled. The event that occurs is chosen finding the first $k$ such that $Q_k > u'$, where $Q_k = \sum_{j=1}^{k}\delta p_j/\delta p$. The state is propagated to be
            \begin{equation}
                \label{eq:MCWF_jump occurs}
                \ket{\widetilde{\psi}(t+\delta t)} = L_k \ket{\psi(t)}.
            \end{equation} 
        \end{enumerate}
    \item The state is normalised $\ket{\widetilde{\psi}(t+\delta t)}\to\ket{\psi(t+\delta t)} = \ket{\widetilde{\psi}(t+\delta t)}/\sqrt{\braket{\widetilde{\psi}(t+\delta t)|\widetilde{\psi}(t+\delta t)}}$.
\end{enumerate}
Note that in this approach no superoperator is assembled and no matrix exponential is calculated. Furthermore, since the trajectories $\Psi_j$ are completely independent of each other, this method can be trivially parallelised by running $N$ trajectories over $N$ different processing nodes to cut down the computational time by a factor of $N$.

\begin{figure}
    \centering
    \includegraphics{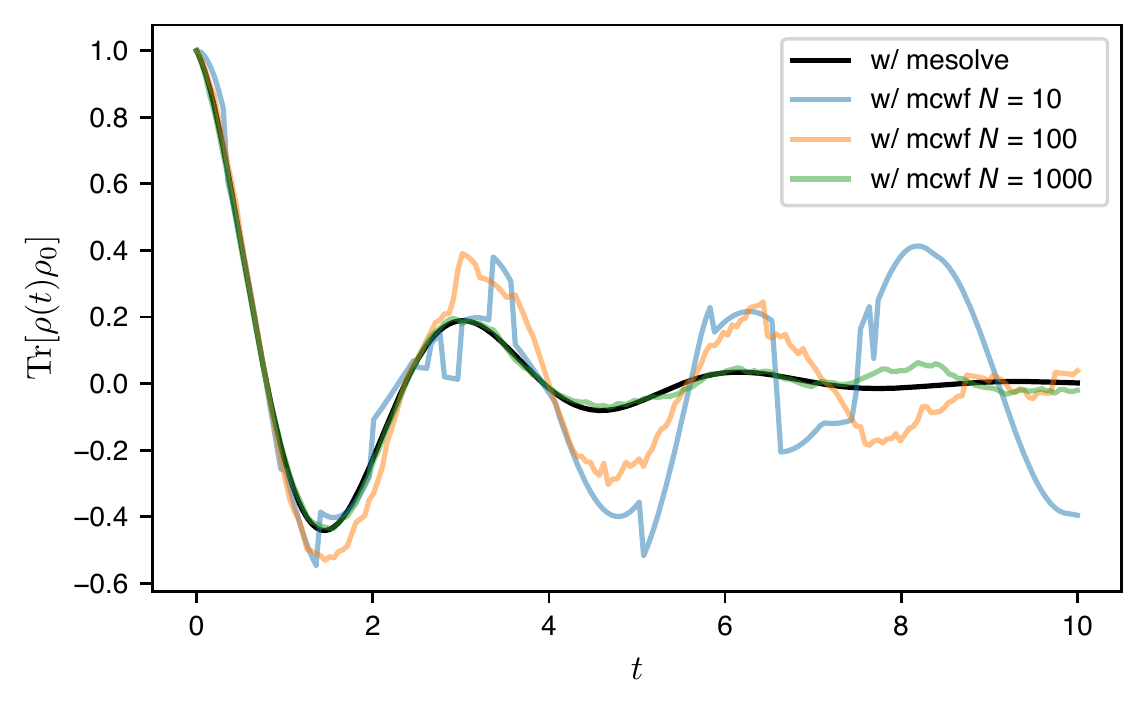}
    \caption{Propagation with stochastic wavefunction method with $N=10,100,1000$ trajectories, using script~\ref{code:mcwf_python} with \texttt{sample} $=N$. The stochastic wavefunction solution approaches the exact one in the limit of large $N$. Here, the solution is compared to that obtained with \texttt{QuTiP}'s finite-difference method \texttt{mesolve}.}
    \label{fig:mcwf}
\end{figure}

A \texttt{python} implementation is presented in the script below, for a two-level system with $H = \sigma_z$ and Lindblad operators $\{ \sigma_z/2, \sigma_x/5\}$, with initial state $\ket{\psi}= (\ket{0}+\ket{1})/\sqrt{2}$ in the $\sigma_z$ basis. The results are shown in Fig.~\ref{fig:mcwf}. Note that the time-step $\delta t$ can be chosen to be a fraction of some operator norm of the Hamiltonian, such that $\delta t \ll \| H\|_\mathrm{op}^{-1}$. An equivalent \texttt{Mathematica} implementation can be found in script~\ref{code:mcwf} in the Appendix. A robust implementation of the stochastic wavefunction method is also available in \texttt{QuTiP}.
\code{https://github.com/frnq/qme/blob/main/python/mcwf.py}{Propagation using stochastic wavefunction method}{code:mcwf_python}{python}{Scripts/python/mcwf.txt}

\subsubsection{Sparse solvers}
\label{ss:sparse_solvers}
When dealing with very large systems, it is worth thinking about sparseness of superoperator and states, since finding its singular value decomposition may become prohibitively expensive.
A number of different techniques can be used to treat sparse and large superoperators, such as
\begin{itemize}
    \item using methods for sparse arrays (\texttt{SparseArray} in \texttt{Mathematica}), such as null-space solvers. A library of linear algebra methods for sparse arrays for \texttt{MATLAB} is available at Ref.~\cite{Davis2006}; 
    \item using Krylov subspace methods to solve for $\mathrm{exp}(\mathcal{L}t)\bm{\rho}_0$ directly~\cite{Vo2017}; Packages \href{https://github.com/weinbe58/expokitpy}{\texttt{expokitpy}} and \href{https://krypy.readthedocs.io/en/latest/#}{\texttt{KryPy}}~\cite{Gaul2014} offer Krylov method implementations for \texttt{python}.
    \item taking the action of the exponential on a given sparse initial state. In \texttt{Mathematica} this can be done with \texttt{MatrixExp} as follows,
    \begin{codeline}{Mathematica}
        \footnotesize
            \begin{minted}[linenos=false]{Mathematica}
                Pt = MatrixExp[M t, psi]
            \end{minted}
    \end{codeline}
    \item using the Arnoldi method~\cite{knap2011emission}. This can be done in \texttt{Mathematica} using the \texttt{Eigensystem} function in combination with \texttt{"Arnoldi"},
        \begin{codeline}{Mathematica}
        \footnotesize
            \begin{minted}[linenos=false]{Mathematica}
                {evals,evecs} = Eigensystem[M t, k, Method -> "Arnoldi"] 
            \end{minted}
        \end{codeline}
    where \texttt{k} represents the index of the eigenvalue (or singular value) to be calculated.
\end{itemize}
However, sometimes the simplest option may be to implement a finite difference method like Runge-Kutta with sparse linear algebra, as it is often just as fast as more sophisticated methods.

\subsection{Correlation functions}
\label{ss:correlation_functions}
Correlation functions measure the relationship between microscopic quantities across time, space and other observables. In statistical mechanics, they are used to calculate the ensemble properties of stochastic processes, and determine the degree of order or randomness in a system. For example, the effect of atmospheric turbulence on the propagation of light beams can be modelled from the correlation functions $C(t,t',\bm{r},\bm{r}') = \braket{n(t,\bm{r})n(t',\bm{r}')}$ of the refractive index $n(t,\bm{r})$~\cite{Paterson2005}. Similarly, the magnetic properties of materials can be inferred from the spatial correlation functions between spins~\cite{Gutzwiller1963}.

In quantum stochastic processes, correlation functions are used to determine the magnitude of decoherence and relaxation processes, as we will discuss in depth in Sec.~\ref{s:bloch-redfield}. The macroscopic properties of a variety of systems can be indeed calculated from the correlation functions of their microscopic features. Of particular importance are, emission and absorption spectra in light-matter interaction (see Sec.~\ref{sss:emiss_abs}), noise power spectra and relaxation rates, bunching and anti-bunching statistics of photons~\cite{Hennrich2005}, electrons~\cite{Emary2012} and other particles. Here, we will examine the basics of correlation functions and show how these can be calculated from the master equation governing the evolution of the density operator. We will then apply these results to calculate the emission spectrum in a simple example of a two-level system interacting with the electromagnetic field. 

\subsubsection{Quantum regression theorem}\label{sss:quantum_regression_theorem}
Linear systems are amply studied in physics because of their simplicity and exact solvability. The equations of motion of the averages of the operators of such systems are often linear, as for the case of Eq.~\ref{eq:superop_form}. 
For these systems, it can be shown that the averages of their two-time correlation functions obey exactly the same equations of motion. This result, first derived by \emph{Lax}, is known as the \emph{quantum regression theorem}~\cite{ficek2005quantum,gardiner2004quantum}, and it provides a method for calculating any two-time correlation function $\braket{A(t)B(t')}$, i.e., involving any two observables at different points in time, for a system whose dynamics are prescribed by a quantum master equation $\dot{\rho} = \mathcal{L}_t[\rho]$~\cite{ficek2005quantum}. 

Suppose that for a certain set of operators $\{A_i\}$, the linear master equation~\eqref{eq:superop_form} yields the following closed system of linear ordinary differential equations to their averages~\cite{Breuer2002},
\begin{equation} \label{Eq:QRT1}
    \frac{d}{d t}\langle A_i(t)\rangle = \sum_j G_{ij}\langle A_j(t) \rangle,
\end{equation}
for some coefficients $G_{ij}$.
Then, their two-point correlation functions
\begin{equation}
    \label{eq:corre_functions}
    \langle A_i(t+\tau)A_l(t)\rangle  = \mathrm{Tr}\big[ A_i \Lambda(t+\tau;t)[A_l\rho(t)]\big],
\end{equation}
where $\Lambda(t;t_0)$ is the dynamical map from time $t_0$ to time $t$, associated with the the master equation $\dot{\rho} = \mathcal{L}_t[\rho]$,
observe the same dynamics, 
\begin{equation} \label{Eq:QRT2}
	\frac{d}{d t}\langle A_i(t+\tau) A_l(t)\rangle = \sum G_{ij} \langle A_j(t+\tau)A_l(t) \rangle.
\end{equation}
Note how the right-hand side of (\ref{eq:corre_functions}) corresponds to the average of $A_i$ at time $t+\tau$ with the choice of initial density operator $\rho\to A_l\rho(t)$~\cite{gardiner2004quantum}.

Any two-time correlation function $\langle A(t+\tau)B(t)\rangle$ can then be simplified using (\ref{eq:corre_functions}) as \cite{Johansson2012},
\begin{align}\label{Eq:TT_corr_simp}
	\langle A(t+\tau)B(t)\rangle &= \mathrm{Tr}[A \Lambda(t+\tau;t)[B\rho(t)]\rbrace], \\ 
 &= \mathrm{Tr}[A \Lambda(t+\tau;t)[B \Lambda(t;0)[  \rho(0)]]].
\end{align}
When calculating $\langle A(t+\tau)B(t)\rangle$ numerically, we can first obtain $\rho(t) = \Lambda(t;0)[\rho(0)]$ with $\rho(0)$ as the initial state. We then propagate $B\rho(t)$ using the dynamical map, to obtain $\Lambda(t + \tau,t)[B\rho(t)]$, and conclude by taking the trace of the resulting operators. If we are interested in steady-state properties, the two-time correlation functions simplify further. By replacing $\rho(0)$ with $\rho(\infty) = \lim_{t\to\infty}\Lambda(t;0)[\rho(0)]$, we can calculate 
$\braket{A(t+\tau)B(t)}$ as
\begin{align}
    \label{eq:steady_state}
    \braket{A(t+\tau)B(t)} &= \tr[A\Lambda(t+\tau;t)[B\rho(\infty)]], \\
    & = \tr[A\Lambda(\tau;0)[B\rho(\infty)]], \\
    & = \braket{A(\tau)B(0)}.
\end{align}

\subsubsection{Emission and absorption spectra}\label{sss:emiss_abs}

Emission and absorption spectra of an optical material can be calculated from the two-time correlation functions of the transition operators associated with the emission and absorption of photons, respectively. For example, an atomic medium given by an ensemble of non-interacting $d$-level systems that interact with the electromagnetic field, will emit light when excited. Its spectrally-resolved intensity is proportional to its emission spectrum $E(\omega)$, which measures the likelihood of transition between eigenstates $\ket{\phi_i}\to\ket{\phi_j}$ with energy difference $\omega$. In first-order perturbation theory, $E(\omega)$ can be calculated using the Fermi golden rule~\cite{Breuer2002}. Line-broadening effects caused by decoherence and relaxation processes can be calculated in second-order perturbation theory using two-point correlation functions and the quantum regression theorem. 

Let us consider a generic two-level emitter with Hamiltonian $H = \Omega\sigma_z/2$ to illustrate how the emission spectrum is calculated. The system can emit a photon via the transition operator $\sigma_- = (\sigma_x-i\sigma_y)/2$ and absorb a photon via its Hermitian conjugate $\sigma_-^\dagger = \sigma_+ = (\sigma_x+i\sigma_y)/2$. Let the system be in a stationary state $\rho(\infty)$. Then, its emission spectrum is calculated from the correlation function of the transition operators $\braket{\sigma_-^\dagger(\tau)\sigma_-(0)}$ as~\cite{Breuer2002},
\begin{align}\label{Eq:TLA_emission_spectrum}
	E(\omega) & \propto \mathcal{F}(\omega)[\braket{\sigma_-^\dagger(\tau)\sigma_-^{\phdagger}(0)}], \\
 & = \int_{-\infty}^{\infty}d\tau e^{-i\omega\tau}\langle{\sigma^\dagger_-}(\tau){\sigma_-^{\phdagger}}(0)\rangle \\ 
 \label{eq:real_form}
 &= 2\;\mathrm{Re}\left\lbrace\int_{0}^{\infty}d\tau e^{-i\omega\tau}\langle{\sigma^\dagger_-}(\tau){\sigma_-^{\phdagger}}(0)\rangle\right\rbrace,
\end{align}
where $\mathcal{F}(\omega)$ is the Fourier transform.
Eq.~(\ref{eq:real_form}) follows from decomposing the limits of the Fourier transform in Eq.~\eqref{Eq:TLA_emission_spectrum} at $t=0$, followed by the use of relation $\langle{\sigma^\dagger_-}(-\tau_+){\sigma_-^{\phdagger}}(0)\rangle  =   \langle{\sigma^\dagger_-}(\tau_+){\sigma_-^{\phdagger}}(0)\rangle^*$, where $\tau_-$ denotes $\tau<0$ and $\tau_+$ denotes $\tau \geq0$~\cite{Breuer2002}.
The generalisation to the emission spectra of a multi-level emitters is obtained by generalisation of Eq.~(\ref{Eq:TLA_emission_spectrum}) as discussed in Ref.~\cite{Hapuarachchi2022}. The emission spectrum $E(\omega)$ is calculated as a sum of all the contributions from the possible transitions $\ket{i}\to\ket{j}$ between the eigenstates of the system with $i>j$, modelled by the operators $\sigma_{ij} = \ketbra{\phi_j}{\phi_i}$,
\begin{equation}
    \label{eq:d-level_emission}
    E(\omega) \propto \sum_{i>j} \mathcal{F}(\omega)[\braket{J_{ij}^\dagger(\tau) J_{ij}^{\phdagger}(0)}],
\end{equation}
with $J_{ij} = \sqrt{\gamma_{ij}}\sigma_{ij}$, for some rates $\gamma_{ij}$.

If the emitter is illuminated by a tunable probe field with angular frequency $\omega_\mathrm{p}$, whose amplitude is assumed to be weak as to not significantly perturb the atom's Hamiltonian, the steady-state probe absorption spectrum can be obtained as follows~\cite{zhou1996absorption, tanas1998analytical, xu2013frontiers},
\begin{equation}\label{Eq:TLA_absorption_spectrum}
	A(\nu) \propto \mathrm{Re}\left\lbrace \int_{0}^{\infty} d\tau e^{i\nu\tau}\langle [\sigma_+^\dagger(\tau),\sigma_+^{\phdagger}(0)]\rangle  \right\rbrace,
\end{equation}
where $\nu = \omega_\mathrm{p}-\omega$ is the detuning of the probe beam relative to the driving laser.

\begin{figure*}[h!]
	\includegraphics[width=\textwidth]{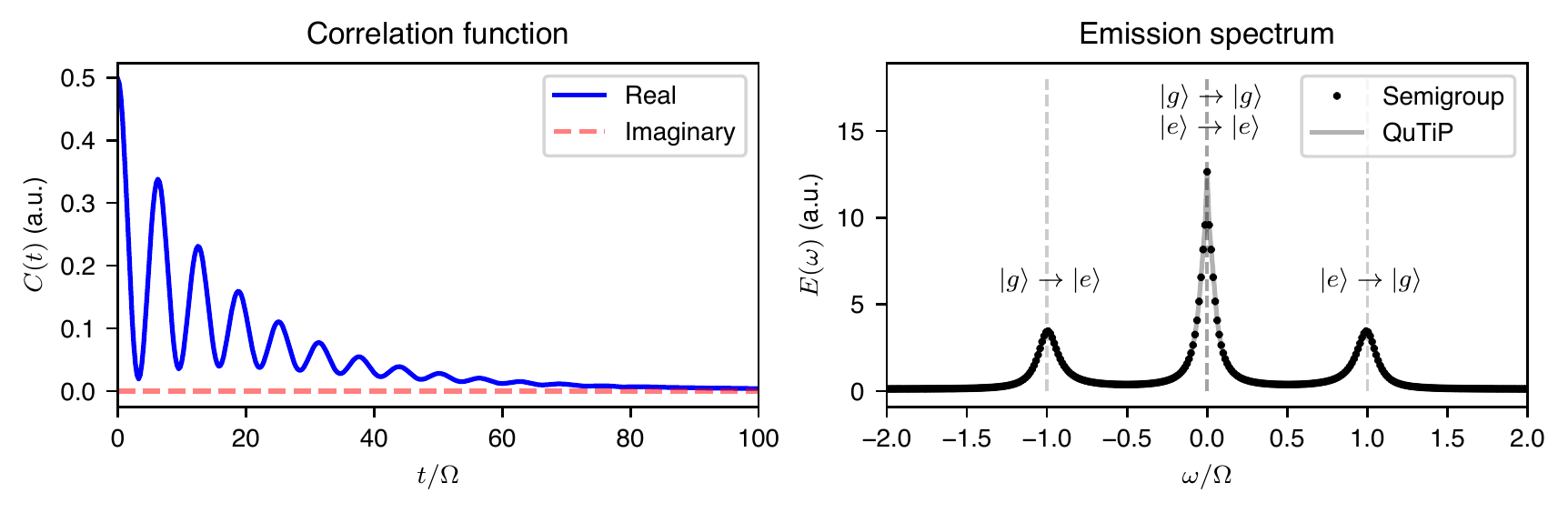}
	\centering
	\caption{ (\textit{Left}) Real and imaginary part of the steady-state correlation function $C(t) = \braket{\sigma_+(\tau)\sigma_-^{\phdagger}(0)}_{ss}$ for a two-level system $H=\Omega\sigma_z/2$, with Rabi frequency $\Omega$ and decay rate $\Gamma = \Omega/10$. (\textit{Right}) Emission spectrum $E(\omega)$ of the considered two-level system associated with transition operator $\sigma_-^{\phdagger}=\ketbra{e}{g}=\sigma_+^\dagger$. The peaks coincide with the transition frequencies $\omega_{ij} = \omega_i - \omega_j$, associated with transitions $\ket{i}\to\ket{j}$ as shown by the labels. The emission spectrum calculated using the semigroup composition rule is compared with the one obtained using \texttt{QuTiP}, using script~\ref{code:emission_qutip}.}
 \label{Fig:Correlation_spectrum}
\end{figure*}

We refer the reader to references \cite{Carmichael1999, meystre2007elements, nation2011qutip} for further details on correlation functions and spectra. We have included the step-by-step implementation of an example of two-level system emission spectrum using (\ref{Eq:TLA_emission_spectrum}) below. The resulting time-domain emission correlation and spectrum are depicted in Fig.\ \ref{Fig:Correlation_spectrum}, alongside the corresponding \texttt{QuTiP} version of the same calculation. 

\code{https://github.com/frnq/qme/blob/main/python/emission_spectrum.py}{Emission spectrum of a two-level atom}{code:emission_spectrum}{python}{Scripts/python/emission_spectrum.txt}

\section{Bloch-Redfield theory}
\label{s:bloch-redfield}

In the previous section we discussed how to implement the Lindblad master equation from a \textit{phenomenological} model of decoherence and relaxation. However, it is sometimes necessary to start from a microscopic description---i.e., the system and environment Hamiltonian---to obtain a master equation for the density operator of the system. When the system interacts weakly with its environment, this can be achieved using Bloch-Redfield theory~\cite{Cohen1992,Breuer2002}. 
This theory is useful when we lack a model for decoherence and relaxation, but we know the nature of the system-environment interactions that drive such processes. As a result, the theory provides a powerful approach to determine the temperature dependence of dephasing and thermalisation rates directly from \textit{first principles}. 

\subsection{Bloch-Redfield master equation}
\label{ss:bloch-redfield_me}

Let us consider a system $\mathrm{S}$, with dimension $\mathrm{dim}_\mathrm{S}$, that interacts with its environment $\mathrm{E}$ according to the following general Hamiltonian
\begin{align}
    \label{eq:BR_microscopic_hamiltonian}
        H &= H_\mathrm{S}+H_\mathrm{E}+H_\mathrm{int} \\
          &= H_\mathrm{S}+H_\mathrm{E} +\sum_\alpha A_\alpha\otimes B_\alpha,
\end{align}
where the coupling operators $A_\alpha$ ($B_\alpha$) are Hermitian and act on the system (environment) such that $H_\mathrm{int}$ is a small perturbation of the unperturbed Hamiltonian $H_0 = H_\mathrm{S}+H_\mathrm{E}$.
Then, under the conditions~\ref{C:BR1}---\ref{C:BR4} discussed in Sec.~\ref{ss:bloch-redfield_approximations}, 
the dynamics of the system's density operator $\rho$ in the eigenbasis $\{\ket{\omega_a}\}$ of $H_\mathrm{S}$~\footnote{$H_\mathrm{S}\ket{\omega_a} = \hbar\omega_a\ket{\omega_a}$.} is prescribed by the Bloch-Redfield master equation,
\begin{eqbox}
    \begin{equation}
        \label{eq:BR_standard_form}
        \dot{\rho}_{ab}(t) = -i\omega_{ab}\rho_{ab}(t)+\sum_{c,d}R_{abcd}\rho_{cd}(t),
    \end{equation}
\end{eqbox}
\noindent
where $\omega_{ab} = \omega_a - \omega_b$ are the frequencies associated with transitions $\ket{\omega_b}\to\ket{\omega_a}$. The Bloch-Redfield tensor $R_{abcd}$ is prescribed by the following expression, where $\delta_{ij}$ is the Kronecker delta,
\begin{equation}
    \label{eq:BR_tensor}
    \begin{split}
    R_{abcd} = -\frac{1}{2\hbar^2}\sum_{\alpha,\beta}\bigg\{ &\delta_{bd}\sum_{n=1}^{\mathrm{dim}_\mathrm{S}} 
    A_{an}^{(\alpha)}A_{nc}^{(\beta)} S_{\alpha\beta}(\omega_{cn}) -  
    A_{ac}^{(\alpha)}A_{db}^{(\beta)} S_{\alpha\beta}(\omega_{ca}) + \\ & \delta_{ac}\sum_{n=1}^{\mathrm{dim}_\mathrm{S}} 
    A_{dn}^{(\alpha)}A_{nb}^{(\beta)} S_{\alpha\beta}(\omega_{dn}) -  
    A_{ac}^{(\alpha)}A_{db}^{(\beta)} S_{\alpha\beta}(\omega_{db})
    \bigg\}.
    \end{split}
\end{equation}
In Eq.~\eqref{eq:BR_tensor}, $A_{ab}^{(\alpha)} = \braket{\omega_a|A_\alpha|\omega_b}$ are the elements of the coupling operators $A_\alpha$ in the eigenbasis of the system Hamiltonian, while $S_{\alpha\beta}(\omega)$ corresponds to the noise-power spectrum of the environment coupling operators~\cite{Jones2017,Gardiner2000},
\begin{equation}
    \label{eq:BR_noise_power_spectra}
    S_{\alpha\beta}(\omega) = \int_{-\infty}^{\infty} d\tau e^{i\omega t}\: \tr\Big[B_\alpha(\tau)B_\beta(0)\rho_\mathrm{E}\Big],
\end{equation}
taken assuming $\rho_\mathrm{E}$ to be some steady state of the environment.

\subsubsection{Thermal relaxation and detailed balance condition}
\label{sss:bloch-redfield_detailed_balance}
When using BR theory it is common to consider environments in thermal equilibrium at inverse temperature $\beta = 1/k_B T$. For example, the environment may be assumed to be in a Bose-Einstein distribution,
\begin{equation}
    \label{eq:BR_gibbs_state}
    G_\beta(H_\mathrm{E}) = \frac{\exp(-\beta H_\mathrm{E})}{\mathcal{Z}},
\end{equation}
with $\mathcal{Z} = \tr[\exp(-\beta H_\mathrm{E})]$, and to be invariant under future evolutions (Gibbs state)~\cite{Breuer2002}.
An out-of-equilibrium density operator that evolves under the dynamics prescribed by Eq.~\eqref{eq:BR_standard_form} with $\rho_\mathrm{E} =G_\beta(H_\mathrm{E})$ will relax towards thermal equilibrium (exchanging energy with the environment). Indeed, the steady state of Eq.~\eqref{eq:BR_standard_form} is itself a Gibbs state $G_\beta(H_\mathrm{S})$ at thermal equilibrium with inverse temperature $\beta$.

The condition for this to occur is known as \textit{detailed balance}, and can be expressed in terms of the ratio between the rates $k_{a\to b}$ associated with transitions $\ket{\omega_a}\to\ket{\omega_b}$ separated by energy $\omega_{ba} = \omega_b -\omega_a$.
\begin{equation}
    \label{eq:BR_detailed_balance}
    \frac{k_{a\to b}}{k_{b\to a}} = \exp(-\beta\omega_{ba}).
\end{equation}
The detailed balance condition implies that the equilibrium populations of the eigenstates of the system follow the Boltzmann distribution $p_a\propto \exp(-\beta \omega_a)$. In terms of noise-power spectra, the detailed balance condition becomes $S_{\alpha\beta}(-\omega)/S_{\alpha\beta}(\omega) = \exp(-\beta\omega)$.

\subsubsection{Example: Spin-boson}
\label{ss:bloch-redfield_spin-boson}
Before discussing the approximation required to derive the BR master equation, let us implement BR theory for the simple and ubiquitous spin-boson model. We consider a two-level system coupled with a large ensemble of \textit{uncorrelated} harmonic oscillators at thermal equilibrium (bosonic bath) 
\begin{equation}
    \label{eq:BR_spin-boson}
    H = \frac{\epsilon_0}{2} \sigma_z+\frac{\Delta}{2} \sigma_z + \sum_k \hbar \omega_k b^\dagger_k b^\phdagger_k + \sigma_z\otimes\sum_k g_k \big(b^\dagger_k + b^\phdagger_k\big),
\end{equation}
where $g_k$ is the strength of the coupling between $\sigma_z$ and some mode $\omega_k$. 

First, we calculate the correlation functions $C_{kk'}(t)$ for the bath operators $B_k = g_k \big(b^\dagger_k + b^\phdagger_k\big)$
\begin{align}
    \label{eq:BR_bath_correlations}
    C_{kk'}(t) &= \delta_{kk'}\tr\big[B_k(t)B_{k'}(0) G_\beta(H_\mathrm{E}) \big], \\
                          &= \frac{g_k^2}{1-\exp(-\beta\omega_{k})}\bigg( e^{-i\omega_k t}+e^{i\omega_k t -\beta\omega_k}\bigg),
\end{align}
where we used the fact that the modes are uncorrelated ($\delta_{kk'}$) and assumed the bath to be in thermal equilibrium  $\rho_\mathrm{E} = G_\beta(H_\mathrm{E})$ at inverse temperature $\beta$, as in Eq.~\eqref{eq:BR_gibbs_state}.

To treat the contribution of a large ensemble of modes, we replace sum over the coupling strength $g_k$ with an integral over some spectral density $J(\omega)$ that well approximates the bath:
\begin{equation}
    \label{eq:spectral_sum_to_integral}
    \sum_k g_k^2 \to \int_0^\infty d\omega J(\omega).
\end{equation}
A common choice is the Ohmic spectral density $J(\omega) = \eta \omega e^{-\omega/\omega_c}$, which is characterised by a cut-off frequency $\omega_c$ and a dimensionless parameter $\eta$, from which we obtain the noise-power~\cite{Lidar2019},
\begin{align}
    \label{eq:BR_NPS}
    S(\omega) &= \int_{-\infty}^{\infty} dt e^{i\omega t} \sum_k C_{kk}(t) \\
              &\approx \int_{-\infty}^{\infty} dt e^{i\omega t} \int_0^{\infty} d\omega' J(\omega') \frac{\big( e^{-i\omega_k t}+e^{i\omega_k t -\beta\omega_k}\big)}{1-\exp(-\beta\omega_{k})} \\
              & = \frac{2\pi\eta\omega \exp(-|\omega|/\omega_c)}{1-\exp(-\beta \omega)}.
\end{align}

We now possess all the elements required to compose the BR tensor of Eq.~\eqref{eq:BR_tensor}. Note that we only have one system coupling operator $A = \sigma_z$, associated with a single noise-power spectrum $S(\omega)$. The following is a \texttt{python} implementation of the Bloch-Redfield tensor, which can then be used to propagate the state of the system using one of the methods discussed in Sec.~\ref{s:doma}. Note that to simplify the solution of Eq.~\eqref{eq:superop_form}, the unitary part of the generator has been absorbed into the tensor $R$,
\begin{equation}
    \label{eq:BR_with_unitary}
    R_{abcd} \to R'_{abcd} =-i\omega_{ac}\delta_{ac}\delta_{bd} + R_{abcd},
\end{equation}
and that system coupling operators are considered to be mutually uncorrelated, $S_{\alpha\beta}=\delta_{\alpha\beta}S_{\alpha\alpha}$.
\code{https://github.com/frnq/qme/blob/main/python/bloch_redfield_tensor.py}{Bloch-Redfield tensor and spin-boson relaxation}{code:BR_spin-boson}{python}{Scripts/python/bloch_redfield_tensor.txt}

\subsection{Approximations for Bloch-Redfield master equation}
\label{ss:bloch-redfield_approximations}
While the Lindblad master equation is guaranteed to be completely positive and trace-preserving\footnote{See Sec.~\ref{s:density_operators} for definition and properties and CPTP maps.}, care must be taken when using BR theory. First, 
the following approximations have to be respected to obtain Eq.~\eqref{eq:BR_standard_form} from the reduced-state von Neumann equation~\cite{Cohen1992,Breuer2002}, as discussed in Sec.~\ref{ss:schrodinger_von_neumann}:

\begin{enumerate}[label=\textbf{C.\arabic*}]
    \item \label{C:BR1} \textbf{Weak coupling approximation:} The interaction $H_\mathrm{int}$ is a small perturbation of the unperturbed Hamiltonian $H_0 = H_\mathrm{S}+H_\mathrm{E}$;
    \item \label{C:BR2} \textbf{Born approximation:} The system-environment density operator is factorised at all times, $\rho_\mathrm{int}(t) = \rho_\mathrm{S}(t)\otimes\rho_\mathrm{E}$, with $\rho_\mathrm{E}$ being some steady state of the environment (justified also by~\ref{C:BR1});
    \item \label{C:BR3} \textbf{Markov approximation:} The bath correlation functions $g_{\alpha\beta}(\tau) = \tr\big[B_\alpha(\tau)B_\beta(0)\rho_\mathrm{E}\big]$ have a short correlation time scale $\tau_\mathrm{E}$, $g_{\alpha\beta}(\tau) \approx 0$ for $\tau \gg \tau_\mathrm{E}$.
    \item \label{C:BR4} \textbf{Rotating wave approximation:} All the contributions from the rapidly oscillating terms, i.e., with characteristic frequency $| \omega_{ab} - \omega_{cd}|\geq \tau^{-1}_\mathrm{E}$, are neglected as they approximately average to zero.
\end{enumerate}

Second, the BR master equation does not, in principle, guarantee positivity of the density operator. That is, when propagating the system in time $\rho(t) = \Lambda_t[\rho_0]$, the populations of $\rho$ may become negative for some time $t>0$~\cite{Whitney2008}. For this reason, when propagating a density operator numerically, it is advisable to check its positivity. The following \texttt{python} script can be used to test positivity, hermitianity and normalisation condition of a density operator. The function \texttt{is\_state(rho)} returns 1 if a \texttt{rho} is a density operator, and a value $s<1$ if \texttt{rho} deviates from the conditions of positivity, hermitianity and normalisation, where $1-s$ is a measure of such deviation.
\code{https://github.com/frnq/qme/blob/main/python/density_operator_test.py}{Is this operator still a state?}{code:density_operator_test}{python}{Scripts/python/density_operator_test.txt}

\subsection{Lindblad form of the Bloch-Redfield master equation}
\label{ss:bloch-redfield_lindblad}
Under certain conditions, it is possible to write the BR master equation in the Lindblad form of Eq.~\eqref{eq:lindblad_master_equation},
\begin{equation}
\label{eq:BR_lindblad_form}
    \dot{\rho}(t) = -\frac{i}{\hbar}[H_\mathrm{S},\rho(t)]+\sum_{\alpha\beta}\sum_{\omega} S_{\alpha\beta}(\omega) \bigg(A_\alpha^\phdagger(\omega)\rho(t) A_\beta^\dagger(\omega) -\frac{1}{2}\Big\{A_\alpha^\dagger(\omega)A_\beta^\phdagger(\omega),\rho(t)\Big\}\bigg),
\end{equation}
where $A_\alpha(\omega) = \sum_{\omega = \omega_b-\omega_a} A_{ab}^{(\alpha)}\ketbra{\omega_a}{\omega_b}$ are the coupling operators in the frequency domain, such that the sum over $\omega$ only needs to be carried out over the transition (Bohr) frequencies $\omega = \omega_b -\omega_a$, as in Eq.~\eqref{eq:BR_tensor}~\cite{Breuer2002}.

This form is useful, for example, to systematically compile the BR tensor from a list of system coupling operators $A_\alpha$ and noise-power spectra $S_{\alpha\alpha}$, or even to compose the full Liouville superoperator associated with the dynamics of Eq.~\eqref{eq:BR_lindblad_form}. 

\subsubsection{Example: Network with random energies and couplings}
\label{ss:bloch-redfield_random_sites}
Let us consider a system consisting of $N$ states $\ket{k}$ with energies $\varepsilon_k$, that interact via couplings $v_{jk}$, with associated Hamiltonian
\begin{equation}
    \label{eq:BR_random_network}
    H_\mathrm{S} = \sum_k \varepsilon_k \ketbra{k}{k} + \sum_{j<k} \bigg(v_{jk}\ketbra{j}{k}+h.c.\bigg).
\end{equation}
Let us assume that each state $\ket{k}$ couples with a local environment of uncorrelated bosonic modes characterised by some noise power spectrum $S_{k}(\omega)$. This type of system-environment model is typically used to model the transport of charge carriers (electrons, holes) or coupled electron-hole pairs (\textit{excitons}) in disordered organic semiconductors~\cite{Jang2018}. In the following \texttt{python} script we study the dynamics of an instance of such random quantum network using Bloch-Redfield theory, with the results shown in Fig.~\ref{fig:random_network}. The BR tensor is calculated using the general method introduced in script~\ref{code:BR_spin-boson}, while the propagator is calculated adaptively for different time scales. A robust and efficient method for the calculation of the Bloch-Redfield tensor is implemented in the \texttt{bloch\_redfield\_tensor} function of \texttt{QuTiP}'s module \texttt{bloch\_redfield}.
\code{https://github.com/frnq/qme/blob/main/python/random_quantum_network.py}{Random quantum network \normalfont{\textsf{(requires script~\ref*{code:BR_spin-boson})}}}{code:random_quantum_network}{python}{Scripts/python/random_quantum_network.txt}

\subsection{Computational resources for Bloch-Redfield master equation}
\label{ss:bloch-redfield_resources}

Markovian master equations like Lindblad and Bloch-Redfield are generally numerically inexpensive when compared to methods involving memory kernels or environmental degrees of freedom~\cite{Breuer2002,Strathearn2018}. Nevertheless, as the size of the system increases, solving density operator master equations can become computationally demanding~\cite{Kondov2001}. Therefore, when implementing BR theory numerically it is important to keep track of the required computational resources.

\subsubsection{Memory requirements}
\label{sss:bloch-redfield_memory_requirements}
Let $d = \mathrm{dim}\mathcal{H}_\mathrm{S}$ be the dimension of the Hilbert space associated with system's Hamiltonian $H_\mathrm{S}$. For any density operator master equation, the amount of complex floating point (FP) numbers required to store the density operator scales with $d^2$, with the coherences (off-diagonal elements) taking up the majority of this memory requirement. Analogously, the memory requirements to store the Liouville superoperator associated with Eqs.~\eqref{eq:lindblad_master_equation} and~\eqref{eq:BR_standard_form} scale as $d^4$. When memory becomes an issue, it is possible to use \textit{stochastic wave function} methods to limit the memory scaling to that of the system dimension ($d$) for the state, and that of the Hamiltonian ($d^2$) for the propagation, as discussed in Sec.~\ref{ss:swfm}.

\subsubsection{Operations requirements}
\label{sss:bloch-redfield_operations_requirements}
There are three main computationally demanding tasks encountered when solving any density operator master equation numerically in Liouville space: 
\begin{itemize}
    \item Constructing the generator of the evolution $\mathcal{L}$, associated with $\dot{\bm{\rho}} = \mathcal{L} \bm{\rho}$;
    \item Computing the propagator $P_t = \exp[ \mathcal{L} t]$;
    \item Propagating the state $\bm{\rho}_t = P_t \bm{\rho}_0$.
\end{itemize}
As discussed in Sec.~\ref{ss:solving_dynamics}, there is an array of approaches to reduce the expense of these tasks, depending on the type of problem. \vspace{5pt}

\noindent
\textbf{\textsf{Propagation}} --- Starting from the bottom, propagating the state in Liouville space involves a matrix multiplication $P \bm{\rho}$ between a $d^2$-vector $\bm{\rho} = \mathrm{vec}(\rho)$ and a $d^2\times d^2$ operator $P$. Without any optimisation, the number of floating point operations required scales with $d^4$~\cite{Kondov2001}. \vspace{5pt}

\noindent
\textbf{\textsf{Matrix exponential}} --- The number of operations required to compute the propagator depends on the method used to calculate the exponential of the matrix associated with $\mathcal{L}$. For example, \href{https://scipy.org/}{\texttt{scipy}}'s implementation (\href{https://docs.scipy.org/doc/scipy/reference/generated/scipy.linalg.expm.html}{\texttt{scipy.linalg.expm}}) uses the Pad\'e method to approximate the matrix exponential (see Refs.~\cite{Moler2003,AlMohy2010} for details on the amount of operations required). This is generally a demanding task, for Lindblad and BR master equations alike: Some approaches to mitigate the computational costs associated with this task are discussed in Sec.~\ref{ss:solving_dynamics}.\vspace{5pt}

\noindent
\textbf{\textsf{Redfield tensor}} --- However, when it comes to constructing the generator of the evolution, calculating the Bloch-Redfield tensor $R$ becomes substantially more demanding than the bare Lindblad generator $\mathcal{L}$.
In essence, this is because each system coupling operator $A_\alpha$ may contribute to any of the $d^2$ transitions $\ketbra{\omega_a}{\omega_b}$ in the eigenbasis of $H_\mathrm{S}$. Therefore, when constructing a Redfield tensor from $m$ coupling operators $A_\alpha$ we may need to perform a number of operations that scales with $m^2 \times d^2$. In constrast, to construct a Lindblad superoperator $\mathcal{L}$ from $m$ \textit{jump} operators $L_k$ we only need a number of operations that scales with $m$. See Ref.~\cite{Kondov2001} for further information on the computational resources required for BR theory, and the efficiency of different numerical implementations.

\subsection{Pauli master equation}
\label{ss:bloch-redfield_pauli}
The computational cost of BR master equations reduces dramatically under some special circumstances. When the system's Hamiltonian $H_\mathrm{S}$ is non-degenerate, the equations of motion for the populations $p_a(t)$ of the eigenstates $\ket{\omega_a}$ are closed and decoupled from the equations of motion for the coherences~\cite{Breuer2002}. The result is a system of linear ordinary differential equations to the populations, known as the Pauli master equation (PME):
\begin{equation}
\label{eq:BR_pauli}
    \dot{p}_a(t) = \sum_b \big[W_{ab}p_b(t)-W_{ba}p_a(t)\big],
\end{equation}
where the matrix elements $W_{ab} = \sum_{\alpha\beta}A_{ba}^{(\alpha)}A_{ab}^{(\beta)}S_{\alpha\beta}(\omega_{ba})$ represent the transition rates between eigenstates $a$ and $b$. 

\begin{figure}
    \centering
    \includegraphics{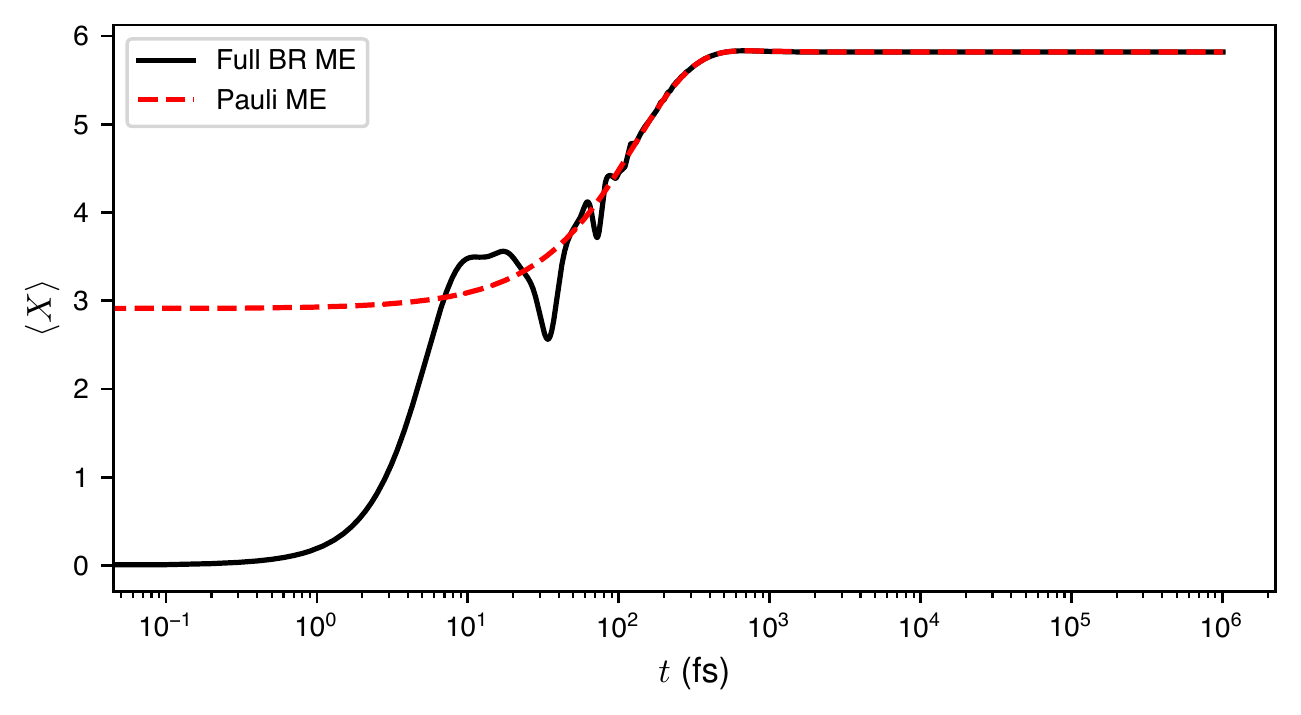}
    \caption{Solution of Bloch-Redfield master equation and associated Pauli master equations for the random quantum network of scripts~\ref{code:random_quantum_network} and~\ref{code:pauli_master_equation}.}
    \label{fig:random_network}
\end{figure}

The Pauli equation~\eqref{eq:BR_pauli} can be written in the vector form $\dot{\bm{p}}(t) = W \bm{p}(t)$ and solved analytically or numerically using the matrix exponential $\bm{p}(t) = \exp[W t]\bm{p}(0)$. Since the population vector $\bm{p}$ is $d$-dimensional, the computational resources required to implement the PME scale with $d^2$. Pauli master equations find applications in scenarios where dephasing happens over a much shorter time scale than thermal relaxation. As an example, room-temperature exciton transport properties have been studied using this approach in Ref.~\cite{Davidson2020,Davidson2022}. The following script implements the PME associated with the problem set up in script~\ref{code:random_quantum_network}. The results are shown in Fig.~\ref{fig:random_network}.

\code{https://github.com/frnq/qme/blob/main/python/pauli_master_equation.py}{Pauli master equation {\normalfont \textsf{(requires script~\ref*{code:random_quantum_network})}}}{code:pauli_master_equation}{python}{Scripts/python/pauli_master_equation.txt}

\section{Periodically driven systems and Floquet theory}
\label{s:oscillatory}

Up until this point, all the Hamiltonians considered are constant, piecewise constant or vary slowly enough that they can be considered piecewise constant. Now we consider the common situation where some part of the Hamiltonian is periodically oscillating in time
\begin{equation}
    \label{eq:periodically_drive}
    H(t) = H_0 + \sin(\omega t + \phi_0) H_1,
\end{equation}
where $H_0$ and $H_1$ are two (generally non-commuting) time-independent Hamiltonians, $\omega$ is some oscillation frequency and $\phi_0\in\mathbb{R}$ is some initial phase. A very common example is a two-level system interacting with an oscillating electric or magnetic field, which is encountered experimentally when driving transitions with a laser or microwave field. However, the approach detailed here is very general and applies to any harmonically oscillating Hamiltonian whose frequency $\omega$ and overtones $k\omega$ ($k\in\mathbb{Z}$) is near resonant with a transition $\ket{E_n}\to\ket{E_m}$ between eigenstates $\ket{E_n}$ with energy $E_n$ of the considered internal Hamiltonian $H_0$,
\begin{equation}
    \label{eq:near-resonant}
    k\:\hbar \omega \approx |E_m - E_n|.
\end{equation}

\subsection{Two-level system interacting with an electric field}
\label{ss:two_level_atom_with_field}

Let's consider a single two-level system (TLS) subjected to an oscillating electric field $\bm{E}(t)$ of wavelength $\lambda$. If the atom is much smaller than $\lambda$, the field would appear spatially constant in the region occupied by the atom. This enables us to write the field as a function of time, 
\begin{equation}
    \label{eq:field}
    \bm{E}(t) = \Big(E_0 e^{-i\omega t} + E_0^* e^{i\omega t}\Big)\hat{z},
\end{equation}
assuming that $\bm{E}$ is oriented along the $\hat{z}$ direction, where $\omega$ is the angular frequency of the incoming radiation.

The total system Hamiltonian $H = H_\mathrm{S} + H_\mathrm{int}$ is the sum of the TLS Hamiltonian,
\begin{equation}
    \label{eq:TLS}
    H_\mathrm{S} = \hbar\frac{\omega_0}{2}\sigma_z,
\end{equation}
with eigenstates $\ket{g}$ and $\ket{e}$,
and the atom-field dipolar interaction Hamiltonian ${H}_\mathrm{int}$~\cite{steck2007quantum, artuso2012optical},
\begin{equation}\label{Eq:TLA_total_Hamiltonian}
H_\text{int} = - \bm{d}\cdot\bm{E},
\end{equation}
where $\bm{d}$ is the transition dipole moment operator of the atom. 
Assuming that the field predominantly interacts with only one electron in the atom, we write $\bm{d}$ in terms of the electron position $\bm{r}_\text{e}$ as $\bm{d} = -e \bm{r}_\text{e}$, where $e$ is the elementary charge. Using a parity argument, it can be shown that the diagonal matrix elements of $\bm{d}$ vanish, i.e., $\langle g|\bm{d}| g\rangle = \langle e|\bm{d}| e\rangle = 0$.
As a result the dipole operator reads
\begin{equation}
    \bm{d} = \braket{e|\bm{d}|g}\ketbra{e}{g} + \braket{e|\bm{d}|g}^*\ketbra{g}{e},
\end{equation}
from which we define the Rabi frequency $\Omega$ of the TLS, and its associated \textit{counter-rotating} frequency $\widetilde{\Omega}$,
\begin{equation}
    \label{Eq:TLA_Rabi}
    \Omega = \braket{g|\bm{d}\cdot \hat{z}|e}\frac{E_0}{\hbar}, \;\;\;\;\widetilde{\Omega} = \braket{e|\bm{d}\cdot \hat{z}|g}\frac{E_0^*}{\hbar}.
\end{equation}
The interaction Hamiltonian then reads 
\begin{equation}
    H_\mathrm{int} = -\hbar\Big(\Omega e^{-i\omega t} + \widetilde{\Omega}e^{i\omega t}\Big)\ketbra{e}{g} -\hbar\Big(\widetilde{\Omega}^* e^{-i\omega t} + \Omega^*e^{i\omega t}\Big)\ketbra{g}{e}.
\end{equation}

\subsubsection{The rotating-wave approximation}
\label{sss:rwa}
Let us now write the full Hamiltonian $H$ in the interaction picture $\widetilde{H} = U_0^\dagger  H U_0$, with $U_0 = \exp(-i H_\mathrm{S} t/\hbar)$,
\begin{equation}
    \label{eq:full_interaction_picture}
        \widetilde{H} = H_\mathrm{S} -\hbar\Big(\Omega e^{-i\Delta \omega t} + \widetilde{\Omega}e^{i(\omega+\omega_0)t}\Big)\ketbra{e}{g} -\hbar\Big(\widetilde{\Omega}^* e^{-i(\omega+\omega_0) t} + \Omega^*e^{i\Delta \omega t}\Big)\ketbra{g}{e}
\end{equation}
with $\Delta \omega = \omega-\omega_0$. If the driving field is close to resonance with the energy splitting of the two-level system, i.e., $\omega\approx\omega_0$, the two time scales involved in the dynamics are separated from each other,
\begin{equation}
    \label{eq:rwa_time-scales}
    \Delta \omega\ll \omega + \omega_0.
\end{equation}
The rapidly oscillating terms in $\omega+\omega_0$, associated with the counter-rotating frequency $\widetilde{\Omega}$, quickly average to zero over the time scale of the Rabi frequency $\Omega$.
As a result the rotating wave approximation (RWA) of the Hamiltonian $H$ in the original frame reads
\begin{equation}
\label{eq:rwa_full}
    H^\mathrm{RWA} = H_\mathrm{S} -\hbar\Big(\Omega e^{-i\omega t}\ketbra{e}{g} + \Omega^* e^{i\omega t}\ketbra{g}{e}\Big).
\end{equation}


\subsubsection{Time-independent Hamiltonian in the rotating frame}
\label{sss:time-independent_rotating_frame}
The Hamiltonian of Eq.~\eqref{eq:rwa_full} can be written in the rotating frame of the driving field, via the transformation generated by the time-dependent unitary $V_\omega = \exp(i H_\mathrm{S} t/\hbar) = \exp(i\omega\sigma_z/2 t)$~ \cite{steck2007quantum},
\begin{equation}\label{Eq:Rotating_frame_transformation}
H^\mathrm{RWA}\to{H}^\mathrm{RWA}_\omega = {V_\omega}{H}{V_\omega}^\dagger + i\hbar\dot{{V}_\omega}{V_\omega}^\dagger.
\end{equation} 
In this frame the Hamiltonian reads
\begin{equation}\label{Eq:H_RF}
\begin{split}
{H}^\mathrm{RWA}_\omega &= \hbar\frac{\Delta \omega}{2}\sigma_z + \hbar\mathrm{Re}[\Omega]\sigma_x+\hbar\mathrm{Im}[\Omega]\sigma_y, \\
    & = \frac{\hbar}{2}\begin{pmatrix}
            \Delta\omega & 2\Omega^* \\
            2\Omega^* & -\Delta\omega 
            \end{pmatrix}.
\end{split}
\end{equation}
This is now a time-independent Hamiltonian in the rotating frame of the driving field, and can be treated with the methods introduced in previous sections. Typically, the decoherence operators are not oscillatory and are also time-independent in this frame, which means solving the master equation also proceeds as above. 

\subsection{Floquet theory and Schr\"odinger evolution}
The RWA is strictly only valid when the Rabi freqency $\Omega$ is small compared to the transition frequency $\omega_0$. When this is not the case, for example in the limit of strong driving inducing multi-photon processes, more sophisticated techniques are required~\cite{Scala2007}.

A common approach to treating strong driving beyond the RWA is using Floquet theory. In this approach, the evolution of a system undergoing periodic variation is expressed in a Fourier series in terms of the oscillation frequency. The Floquet theorem states that a set of time-dependent differential equations whose coefficients vary periodically will have solutions with the same periodicity. This is the temporal equivalent of Bloch's theorem in space, with the solution expressed in terms of quasi-energies instead of quasi-momenta.

In the context of quantum systems, Floquet theory provides a method for finding solutions to the time-dependent Schr\"odinger equation due to the influence of a time-periodic Hamiltonian. The Floquet treatment of the two-level system problem under strong driving was treated by Shirley~\cite{Shirley1965}. However, the approach is of general validity and invaluable in a variety of time-dependent problems, such as analogue quantum simulation~\cite{Kyriienko2018}, quantum information processing~\cite{Bomantara2018}, heat engines and laser cooling~\cite{Restrepo2018}, quantum optimal control~\cite{Bartels2013,Castro2022}, and time crystals~\cite{Else2016,Sacha2018}.

\subsubsection{Floquet modes and quasi-energies}
\label{sss:floquet_modes_quasi_energies}

Let us consider the time-dependent Schr\"odinger equation for a periodic Hamiltonian $H(t) = H(t+n T)$, for all $n\in\mathbb{Z}$,
\begin{equation}
\label{eq:shroedinger_time-dep}
    i \hbar \frac{d}{dt} \ket{\psi(t)} = H(t) \ket{\psi(t)}.
\end{equation}
The Floquet theorem states that the general solution has the form
\begin{equation}
    \label{eq:floquet_theorem}
    \ket{\psi(t)} = \sum_{\alpha}e^{-i\epsilon_\alpha t/\hbar}\ket{\phi_\alpha(t)},
\end{equation}
where $\ket{\phi_\alpha(t)} = \ket{\phi_\alpha(t+n T)}$ are some periodic functions, known as \textit{Floquet modes}, and $\epsilon_\alpha$ are the associated quasi-energies, constant in time and uniquely defined up to multiples of $\omega = 2\pi/T$~\cite{Shirley1965}. By plugging Eq.~\eqref{eq:floquet_theorem} back into Eq.~\eqref{eq:shroedinger_time-dep}, we can recast the problem as an eigenvalue problem to the quasi-energies for the operator $\mathrm{H}(t):=H(t) - i\hbar d_t$,
\begin{equation}
    \label{eq:quasi_energy_eigen}
    \mathrm{H}(t) \ket{\phi_\alpha(t)} = \epsilon_\alpha\ket{\phi_\alpha(t)}.
\end{equation}
This equation can be solved numerically or analytically in order to find the quasi-energies and the Floquet modes. An alternative approach to finding the solution is to solve the eigenvalue problem posed by the propagator $U(t+nT;t)$~\cite{Creffield2003},
\begin{equation}
    \label{eq:alternative_floquet_solution}
    U(t+nT;t)\ket{\phi_\alpha (t)} = e^{-i\epsilon_\alpha T/\hbar}\ket{\phi_\alpha(t)},
\end{equation}
with is then solved for $\eta_\alpha = \exp(-i\varepsilon_\alpha T/\hbar)$, to find $\varepsilon_\alpha = -\hbar \arg(\eta_\alpha)/T$. This approach is implemented in \texttt{QuTiP} with the \texttt{floquet\_modes} method.

\subsubsection{The Floquet Hamiltonian and Fourier analysis}
\label{sss:floquet_fourier}

Thanks to their shared periodicity we can express both the Hamiltonian and the Floquet modes as Fourier series,
\begin{equation}
    \ket{\phi_\alpha(t)} = \sum_n e^{-i \omega n t} \ket{\alpha,n}, \quad  H(t) = \sum_n e^{-i \omega n t} H_n,
\end{equation}
where we have implicitly introduced the Fourier components $\ket{\alpha,n}$ and $H_n$ of the Floquet modes and of the Hamiltonian, respectively, 
\begin{equation}
    \label{eq:Floquet_fourier_components}
    \ket{\alpha,n} = \frac{1}{T}\int_0^T dt e^{i\omega n t} \ket{\phi_\alpha(t)}, \quad
    H_n = \frac{1}{T}\int_0^T dt e^{i \omega n t} H(t).
\end{equation} 
This allows us to define a \textit{Floquet Hamiltonian}, $H_F$, whose components are given by
\begin{equation}
    \bra{\alpha, n} H_F \ket{\beta, m} = H_{n-m}^{(\alpha,\beta)} + n \omega \delta_{\alpha \beta} \delta_{n m},
\end{equation}
which can be used to calculate transition probabilities $P_{\alpha\to\beta}(t)$ between the modes $\alpha\to\beta$, as discussed in the next section.

\subsubsection{Transition probabilities from Floquet Theory}
\label{sss:Floquet_transition_probabilities}

Let us consider a simple sinusoidal variation in the Hamiltonian, such that $H$ has a finite Fourier series
\begin{align}
\label{eq:example_sinusoidal}
    H(t) &= \sum_{n=-1}^{1} e^{-i\omega n t} H_n, \\
         &= H_0 + \widetilde{H}_1 \cos(\omega t),
\end{align}
with $\widetilde{H}_1 := H_{-1} e^{-i\omega t} + H_1 e^{i\omega t}$. Then, the Floquet Hamiltonian has the general structure
\begin{equation}
    H_F = \left( \begin{array}{ccccc}
    H_0 - 2 \hbar \omega & H_1 & 0 & 0 & 0 \\
    H_{-1} & H_0 - \hbar \omega & H_1 & 0 & 0 \\
    0 & H_{-1} & H_0 & H_1 & 0 \\
    0 & 0 & H_{-1} & H_0 + \hbar \omega & H_1 \\
    0 & 0 & 0 & H_{-1} & H_0 + 2 \hbar \omega \\
    \end{array}
    \right)
\end{equation}
where the size of the matrix is limited by the number of harmonics included in the Fourier expansion. If we then diagonalise $H_F$, the time dependent wavefunction can be written in terms of the eigenvectors $\ket{\lambda}$ and corresponding eigenvalues $\lambda$ of the Floquet Hamiltonian
\begin{equation}
    \label{eq:floquet_eigenvalue}
    H_F \ket{\lambda} = \lambda \ket{\lambda}.
\end{equation}
The time-dependent wavefunction $\ket{\psi(t)} = U(t;t_0)\ket{\psi(t_0)}$ is then expressed in terms of the propagator $U(t;t_0)$, whose elements can be written as
\begin{equation}
    U_{\beta \alpha}(t;t_0) = \sum_n \bra{\beta, n}\exp[-i H_F (t-t_0)/\hbar]\ket{\alpha, 0} e^{i n \omega t} = \sum_n \sum_{\lambda} \braket{\beta, n|\lambda}\braket{\lambda|\alpha, 0} e^{-i\lambda (t-t_0)/\hbar} e^{i n \omega t}.
\end{equation}
The probability at time $t$ of a given transition $\alpha \rightarrow \beta$ between Floquet modes with quasi-energies $\epsilon_\alpha$, $\epsilon_\beta$ can then be computed directly,
\begin{equation}
    P_{\alpha \rightarrow \beta}(t-t_0) = \sum_k|\bra{\beta k}\exp[-i H_F (t-t_0)/\hbar]\ket{\alpha 0}|^2.
\end{equation}

\begin{figure}
    \centering
    \includegraphics[width=0.80\textwidth]{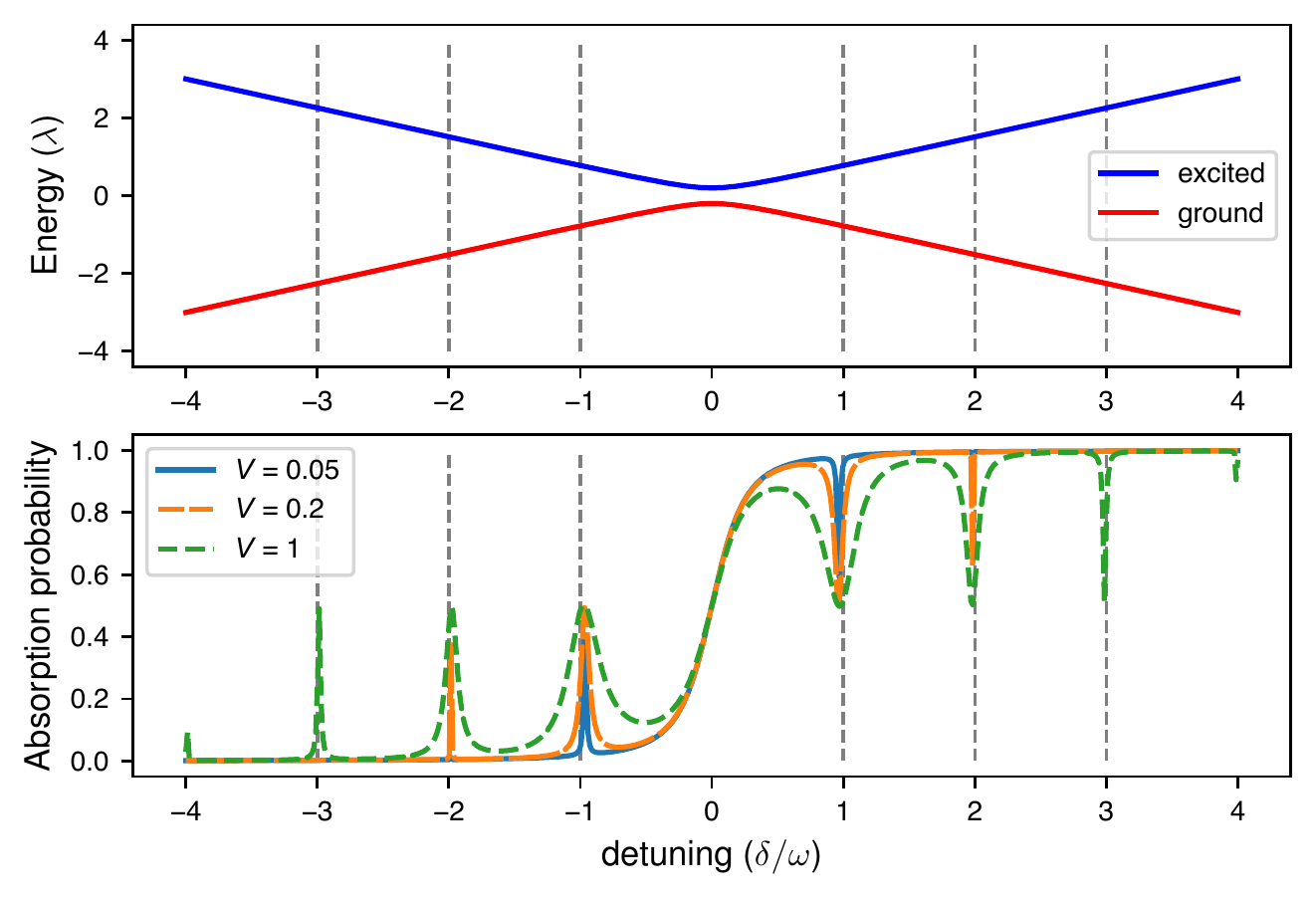}
    \caption{Absorption probability associated with the transition $\ket{g}\to \ket{e}$ in a two level system with Hamiltonian $H_0 = \delta\sigma_z/2+\varepsilon\sigma_x$ and eigenvalues $\lambda_g,\lambda_e$. The system is strongly driven via the interaction $H_{int} = V\sigma_z/2$ with a cavity mode of frequency $\omega$. The vertical dashed lines correspond to the $n$-photon transitions, which are enabled as the interaction strength $V$ increases. The figures is generated using script~\ref{code:floquet_plot_python}.}
    \label{fig:floquet}
\end{figure}

In addition, because the time evolution is given by the Floquet components, the time-averaged probability $\overline{P}_{\alpha\to\beta}$ can be evaluated as
\begin{equation}
\label{eq:time-averaged}
    \overline{P}_{\alpha \rightarrow \beta} = \sum_k \sum_{\lambda}|\braket{\beta k|\lambda}\braket{\lambda|\alpha 0}|^2.    
\end{equation}
This equation is implemented in the following \texttt{python} script for a system given by a two-level system interacting with a quantised electromagnetic field mode $a^\dagger$ with frequency $\omega$,
\begin{equation}
    \label{eq:floquet_example}
    H = H_\mathrm{S} + V\sigma_z (a^\dagger e^{-i\omega t}+a e^{i\omega t}) + \hbar \omega a^\dagger a,
\end{equation}
under different driving strengths $V$, as shown in Fig.~\ref{fig:floquet}. The size of the Floquet Hamiltonian scales with both the number of states and the number of modes included in the Floquet expansion. The relative magnitude of $||H_1||$ to $||H_0||$ controls how many modes need to be included. In practice, this can be determined by increasing the number of modes until the result converges. It is worth noting that this method can be computationally costly due to the size of the Floquet Hamiltonian. However, if convergence can be achieved, the method is \textit{exact} and therefore can be used to compute the effects of strong driving, multi-photon transitions and other effects beyond the rotating wave approximation.
\code{https://github.com/frnq/qme/blob/81c4f16e45c7cf80ddee873814ee9d27f13d325d/python/floquet_rates_python.py}{Transition probability with Floquet theory}{code:floquet_wave_vector_python}{python}{Scripts/python/floquet_rates.txt}

\subsubsection{Extension of Floquet theory to decoherence processes}
\label{sss:extension_to_decoherence}

While the extension of Shirley's approach to model decoherence is less well established, there have been a number of different approaches, depending on how the expansion in Floquet components is introduced to the master equation~\cite{Breuer2000,Wu2010,Schnell2020,Mori2023} as well as other approaches to including beyond-rotating wave physics into a master equation treatment~\cite{Scala2007,Werlang2008,Majenz2013,Muller2017,Kohler2018}.

One approach, which is also relatively simple to code, was introduced by Bain and Dumont~\cite{Bain2010} to model higher order corrections in magic angle spinning NMR experiments. In their approach they use the Liouville form introduced in section~\ref{ss:superoperator}, and express a periodic superoperator $\mathcal{L}_t$ as a Floquet expansion, resulting in a \textit{Floquet superoperator} $\mathcal{L}_F$ that generates the dynamics in an effective time-independent Markovian master equation, in analogy with the Floquet Hamiltonian in the Shirley approach. However, it is important to notice the existence of a time-independent Floquet superoperator $\mathcal{L}_F$ is not always guaranteed, as shown in Ref.~\cite{Schnell2020}. In fact, depending on the choice of $\mathcal{L}_t$, the evolution might be described by an equivalent non-Markovian master equation that is homogeneous in time but not time-local. Although more computationally demanding than the standard Floquet approach, this extension to decoherence processes is quite general and can be applied to master equations with oscillatory Hamiltonian components fairly easily~\cite{Bushev2010,Schoen:2020}.

\section{Discussion}
\label{s:discussion}

In this tutorial we have covered the basics of quantum master equations, showcasing their significance with examples and discussions. The methods reviewed here, such as the GKSL master equation and Bloch-Redfield theory are the cornerstone of stochastic quantum dynamics, and constitute only a small fraction of the developed field of open quantum systems. For further readings on these topics we direct the authors to the following textbooks~\cite{Breuer2002,Nielsen2010,Gardiner2000,gardiner2004quantum,Stenholm2005,Benatti2003,Schlosshauer2007,Bengtsson2006,Rivas2012} and reviews~\cite{Tanimura2006,Rotter2015,Breuer2016,Milz2017}. The power of quantum master equations goes well beyond the considered systems and examples. The theory has been extended to non-Markovian dynamics~\cite{Yu1999,Breuer1999,Yu2004,Ferialdi2016,Piilo2008,Zhang2019b}, non-linear systems~\cite{Kilin1986,Mancal2012}, time-convolutionless master equations~\cite{Pereverzev2006,Nan2009,Timm2011,Kidon2015}, and is in constant development~\cite{Yang2016a,Dominikus2021,Ferguson2021,Wu2022,Davidovic2022,Donvil2022}. 

Further research in this field has been focusing on several aspects, such as extending the applicability of QMEs beyond the standard approximations~\cite{Stenius1996,Whitney2008,Davidovic2020}, the combination of QMEs with compression methods~\cite{Cygorek2022} such as tensor networks~\cite{Werner2016,Xu2019,Jorgensen2019,Fugger2020,Nakano2021}, the use of neural networks~\cite{Hartmann2019,Liu2022}, and the quantum simulation of open system dynamics~\cite{DiCandia2015,Endo2020,Schlimgen2022,Kamakari2022}. These exciting developments are set to expand the range of applicability of QMEs to problems that are typically hard to solve, such as the dynamics of correlated many-body quantum systems that underlie the physics of quantum phase transitions~\cite{DeChiara2018,Heyl2019,Bayha2020,Carollo2020,Rossini2021}, quantum computing architectures~\cite{Wu2021,Brown2021,Bravyi2022,Madsen2022}, optoelectronic devices~\cite{Chenu2015,Scholes2020a}, and complex chemical reactions~\cite{Cao2019,Schroder2019,Mcardle2020,Ye2021}.

\section*{Acknowledgments}
\label{s:acknowledgments}
\noindent
The Authors acknowledge the Australian Research Council (grant number CE170100026) for funding and the National Computational Infrastructure (NCI), supported by the Australian Government, for the computatioal resources. HH gratefully acknowledges Dinuka U Kudavithana for insightful discussions. FC acknowledges that results incorporated in this standard have received funding from the European Union Horizon Europe research and innovation programme under the Marie Sklodowska-Curie Action for the project SpinSC. JHC wishes to thank A. Greentree, J. Ang, S. Andr\'{e}, C. M\"{u}ller, J. Jeske, N. Vogt and several other collaborators for useful input and corrections over the 15 years we used the set of technical notes on superoperators that were the inspiration for this tutorial.


\newpage
\clearpage

\appendix

\noindent
{\bfseries\sffamily\LARGE{Appendix}}

\renewcommand{\theequation}{B-\arabic{equation}}
\setcounter{equation}{0}  

\section{Examples in Mathematica, MATLAB and QuTiP}
\label{a:examples}

\code{https://github.com/frnq/qme/blob/main/python/qutip/tensor_partial.py}{Tensor product and partial trace using QuTiP}{code:tensor_partial}{python}{Scripts/python/qutip/tensor_partial.txt}

The following script implements a symbolic steady-state solution using \texttt{MATLAB}. 
\code{https://github.com/frnq/qme/blob/main/matlab/symbolic_solution.m}{Symbolic steady-state solution}{code:symbolic_matrix_solving}{MATLAB}{Scripts/matlab/symbolic_solution.txt}

Alternatively, we could solve the set of coupled algebraic equations obtained by element-wise comparison of the left and right hand sides of Eq.~\eqref{eq:lindblad_master_equation}, by replacing the last two code lines above with the following.
\code{https://github.com/frnq/qme/blob/main/matlab/steady_state_algebraic.m}{Solving element-wise algebraic equations {\normalfont \textsf{(requires script~\ref*{code:symbolic_matrix_solving})}} }{code:element-wise_algebraic}{MATLAB}{Scripts/matlab/steady_state_algebraic.txt}

\code{https://github.com/frnq/qme/blob/main/matlab/null_space_matlab.m}{Solving for the null space of superoperator}{code:null_space_MATLAB}{MATLAB}{Scripts/matlab/null_space.txt}

\code{https://github.com/frnq/qme/blob/main/python/qutip/mesolve_example.py}{Solution using QuTiP {\normalfont\textsf{(requires scripts~\ref*{code:super_matrix_exp})}}}{code:mesolve_example}{python}{Scripts/python/qutip/mesolve_example.txt}

\code{https://github.com/frnq/qme/blob/main/matlab/dynamics_norm_svd.m}{Solution using normalized singular vectors}{code:temporal_solution}{MATLAB}{Scripts/matlab/dynamics_norm_svd.txt}

Alternatively, we could obtain the same solution using non-normalized singular vectors as
\begin{equation}
    \label{eq:non_normalised_svd_solution}
    \bm{\rho}(t) = \sum^{d^2}_{k=1}b_k\bm{R}_k e^{\lambda_k t}
\end{equation}
where the coefficients $b_k$ are found by performing row reduction on the following augmented matrix formed with $\bm{R}_k$'s and $\bm{\rho}(0)$ as columns,
\begin{equation}\label{Eq:Row_reduction_mat}
    R = \left(\;\bm{R}_1 \;\; \bm{R}_2 \;\; \hdots \;\; \bm{R}_{d^2} \;\big|\; \bm{\rho}(0)\; \right).
\end{equation}
When the above matrix is in row echelon form, the right hand column will give the values of $b_k$'s. The following continuation of the earlier code implements the alternative method and the resulting excited state population.
\code{https://github.com/frnq/qme/blob/main/matlab/dynamics_non_norm_svd.m}{Solution using non-normalized singular vectors  {\normalfont \textsf{(requires script~\ref*{code:temporal_solution})}}}{code:non-normalised}{MATLAB}{Scripts/matlab/dynamics_non_norm_svd.txt}

\code{https://github.com/frnq/qme/blob/main/matlab/dynamics_finite_diff.m}{Solution with finite-difference methods {\normalfont \textsf{(requires script~\ref*{code:temporal_solution})}}}{code:solution_matlab_ode}{MATLAB}{Scripts/matlab/dynamics_finite_diff.txt}

\begin{figure}  
	\includegraphics[width=0.6\columnwidth]{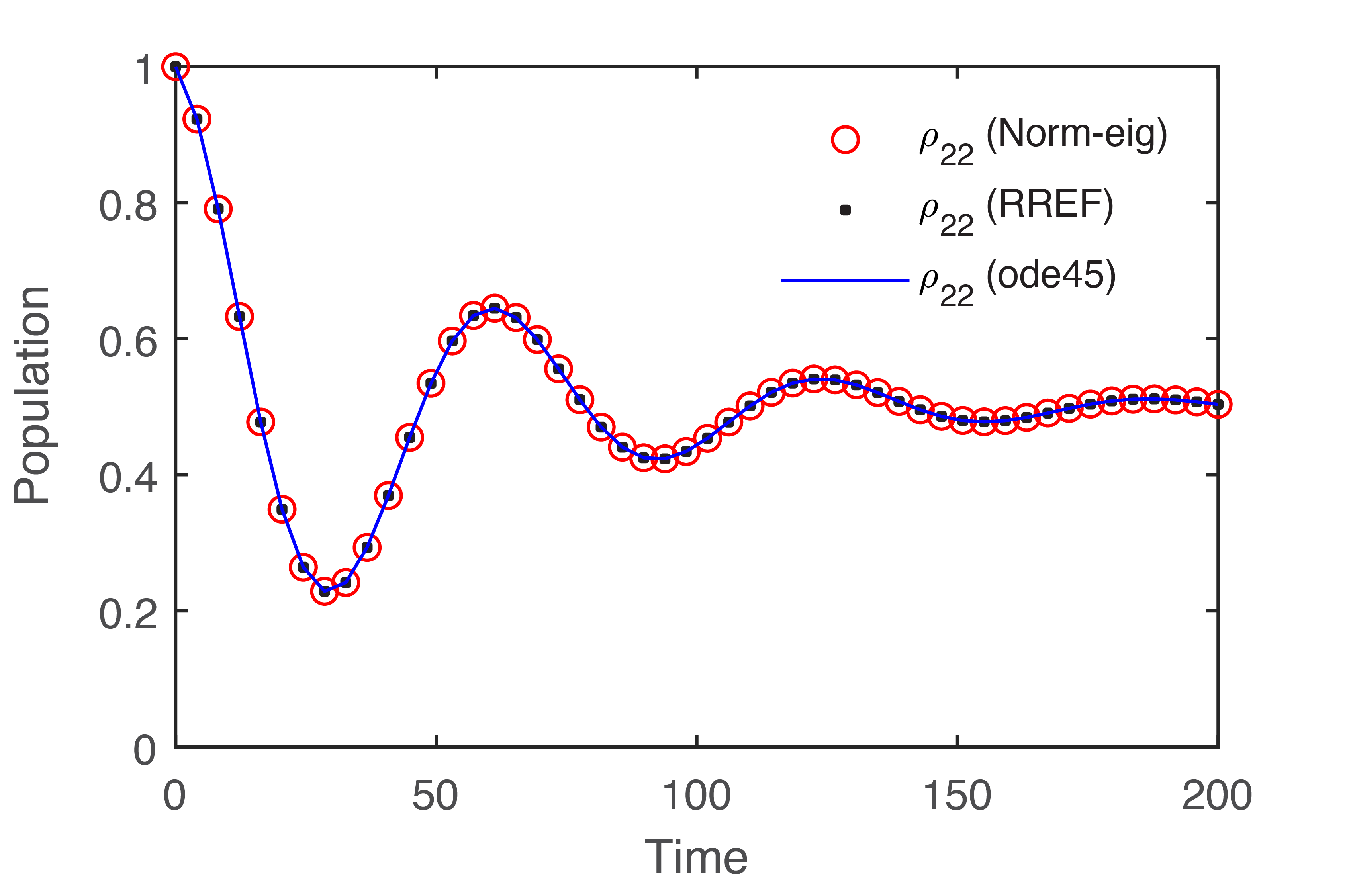}
	\centering
	\caption{Temporal evolution of the excited state population ($\rho_{22}$) obtained using the normalized superoperator eigenvectors (Norm-eig), non-normalized eigenvector (reduced) row-echelon-form (RREF), and numerical differential equation solving (ode45).\label{Fig:Temporal_sol}} 
\end{figure}

\code{https://github.com/frnq/qme/blob/main/mathematica/mcwf.nb}{Propagation using stochastic wavefunction method}{code:mcwf}{Mathematica}{Scripts/mathematica/mcwf.txt}

\code{https://github.com/frnq/qme/blob/main/python/qutip/emission_spectrum.py}{Emission spectrum using QuTiP {\normalfont\textsf{(requires scripts~\ref*{code:superoperator_python} and~\ref*{code:emission_spectrum})}}}{code:emission_qutip}{python}{Scripts/python/qutip/emission_spectrum.txt}

The following script~\ref{code:floquet_plot_python} can be used to obtain the plots shown in Fig.~\ref{fig:floquet}.
\code{https://github.com/frnq/qme/blob/main/python/floquet_plot.py}{Transition probability of two-level atom under strong driving}{code:floquet_plot_python}{python}{Scripts/python/floquet_plot.txt}

\code{https://github.com/frnq/qme/blob/main/matlab/floquet_rates.m}{Function for the time-evolution of a two-level atom under strong driving}{code:floquet_wave_vector}{MATLAB}{Scripts/matlab/floquet_rates.txt}

\code{https://github.com/frnq/qme/blob/main/matlab/floquet_plot.m}{Time-evolution of a two-level atom under strong driving}{code:floquet_wave_vector_plot}{MATLAB}{Scripts/matlab/floquet_plot.txt}

\section{Software requirements}
\label{a:software_requirements}

The \texttt{python} scripts in the \textit{main text} have been tested using \texttt{Python 3.9.6}, and require the libraries \texttt{numpy}, \texttt{scipy} and \texttt{matplotlib}. The \texttt{python} scripts in the \textit{appendix} also require the libraries \texttt{qutip}, \texttt{sympy} and \texttt{tqdm}. The \texttt{Mathematica} (\texttt{MATLAB}) scripts in the appendix were tested using version \texttt{13.2} (\texttt{R2022a}), and do not require any additional library.

\end{document}